\documentclass[12pt]{article}
\usepackage{graphicx}
\usepackage{epsf}
\usepackage{psfig}
\usepackage{a4}  
\def\hybrid{\topmargin 0pt      \oddsidemargin 0pt
        \headheight 0pt \headsep 0pt
        \textheight 9in         
        \textwidth 6.25in       
        \marginparwidth .875in
        \parskip 5pt plus 1pt   \jot = 1.5ex}

\catcode`\@=11
\def\marginnote#1{}
\newcount\hour
\newcount\minute
\newtoks\amorpm
\hour=\time\divide\hour by60 \minute=\time{\multiply\hour by60
\global\advance\minute by-\hour}
\edef\standardtime{{\ifnum\hour<12 \global\amorpm={am}%
        \else\global\amorpm={pm}\advance\hour by-12 \fi
        \ifnum\hour=0 \hour=12 \fi
        \number\hour:\ifnum\minute<10 0\fi\number\minute\the\amorpm}}
\edef\militarytime{\number\hour:\ifnum\minute<10
0\fi\number\minute}

\def\draftlabel#1{{\@bsphack\if@filesw {\let\thepage\relax
   \xdef\@gtempa{\write\@auxout{\string
      \newlabel{#1}{{\@currentlabel}{\thepage}}}}}\@gtempa
   \if@nobreak \ifvmode\nobreak\fi\fi\fi\@esphack}
        \gdef\@eqnlabel{#1}}
\def\@eqnlabel{}
\def\@vacuum{}
\def\draftmarginnote#1{\marginpar{\raggedright\scriptsize\tt#1}}

\def\draft{\oddsidemargin -.5truein
        \def\@oddfoot{\sl preliminary draft \hfil
        \rm\thepage\hfil\sl\today\quad\militarytime}
        \let\@evenfoot\@oddfoot \overfullrule 3pt
        \let\label=\draftlabel
        \let\marginnote=\draftmarginnote
   \def\@eqnnum{(\theequation)\rlap{\kern\marginparsep\tt\@eqnlabel}%
\global\let\@eqnlabel\@vacuum}  }

\def\numberbysection{\@addtoreset{equation}{section}
        \def\theequation{\thesection.\arabic{equation}}}

\def\underline#1{\relax\ifmmode\@@underline#1\else
        $\@@underline{\hbox{#1}}$\relax\fi}

\def\titlepage{\@restonecolfalse\if@twocolumn\@restonecoltrue\onecolumn
     \else \newpage \fi \thispagestyle{empty}\c@page\z@
        \def\thefootnote{\fnsymbol{footnote}} }

\def\endtitlepage{\if@restonecol\twocolumn \else  \fi
        \def\thefootnote{\arabic{footnote}}
        \setcounter{footnote}{0}}  
\catcode`@=12 \relax

\def\beq{\begin{equation}}
\def\eeq{\end{equation}}
\def\bea{\begin{eqnarray}}
\def\eea{\end{eqnarray}}
\def\nn{\nonumber}

\def\neq{\not=}
\relax 

\numberbysection \hybrid
\begin{document}
\begin{titlepage}
\begin{center}
April~2001 \hfill    PAR--LPTHE 01/15 \\[.5in]
{\large\bf Models $WD_{n}$ in the presence of disorder
and the coupled models.}\\[.5in]
        \bf Vladimir S. Dotsenko, Xuan Son Nguyen and Raoul Santachiara \\
        {\it LPTHE\/}\footnote{Laboratoire associ\'e No. 280 au CNRS}\\
       \it  Universit\'e Pierre et Marie Curie, PARIS VI\\
        Boite 126, Tour 16, 1$^{\it er}$ \'etage \\
        4 place Jussieu\\
        F-75252 Paris CEDEX 05, FRANCE\\
        dotsenko@lpthe.jussieu.fr, xuanson@lpthe.jussieu.fr,
santachi@lpthe.jussieu.fr
\end{center}
\end{titlepage}
\underline{Abstract.}

We have studied the conformal models $WD_{n}^{(p)},
\,\,\,n=3,4,5,...,$ in the presence of disorder which couples to
the energy operator of the model. In the limit of $p\gg 1,$ where
$p$ is the corresponding minimal model index, the problem could be
analyzed by means of the perturbative renormalization group, with
$\epsilon$ - expansion in $\epsilon=\frac{1}{p}$. We have found
that the disorder makes to flow the model $WD_{n}^{(p)}$ to the
model $WD_{n}^{(p-1)}$ without disorder. In the related problem of
$N$ coupled regular $WD_{n}^{(p)}$ models (no disorder), coupled
by their energy operators, we find a flow to the fixed point of
$N$ decoupled $WD_{n}^{(p-1)}$. But in addition we find in this
case two new fixed points which could be reached by a fine tuning
of the initial values of the couplings. The corresponding critical
theories realize the permutational symmetry in a non-trivial way,
like this is known to be the case for coupled Potts models, and
they could not be identified with the presently known conformal
models.

\newpage

\section{Introduction}

With the present knowledge of the effect of disorder on the
critical behavior of statistical systems it could be fair to say
that any extra theoretical model will be interesting in which
these effects are nontrivial and can be studied reliably.

In this respect we find the set of conformal theory models
$WD_{n}^{(p)},\,\,\,\, n=3,4,5,...,$ to be interesting. Here $p$
is the index of the corresponding unitary series of minimal
models. These models are based on the corresponding $W$ - chiral
algebras. They have been defined in the paper \cite{Fateev}.

The simplest model in this series $WD_{3}^{(p)}$ will be
sufficient to give a preliminary discussion of our approach and
results.

The model itself, which has not much been used in statistical
physics applications, will be described in the next Section. For
the moment we shall observe that a particular primary operator in
this theory, which is naturally identified with the energy
operator, $\varepsilon(x)$, possesses conformal dimension
\begin{equation}
\Delta_{\varepsilon}=\frac{1}{2}
\end{equation}
in the limiting theory $WD_{3}^{(p)}, p\rightarrow\infty $.
Physical dimension of $\varepsilon $ will be twice bigger,
$\Delta_{\varepsilon}^{ph}=2\Delta_{\varepsilon}=1$. For $p$
finite, $\Delta_{\varepsilon}$ is of the form:
\begin{equation}
\Delta_{\varepsilon}=\frac{1}{2}-\frac{5}{2}\epsilon
\end{equation}
where
\begin{equation}
\epsilon=\frac{1}{p+1}
\end{equation}
If we couple disorder to $\varepsilon(x)$:
\begin{equation}
\int d^{2}x\quad m(x)\varepsilon(x)
\end{equation}
with $m(x)$ being random valued, then the disorder produced
interaction of the replicated theory:
\begin{equation}
g\int d^{2}x\sum^{N}_{a\neq
b}\varepsilon_{a}(x)\varepsilon_{b}(x)
\end{equation}
will be marginal in the limit $p\rightarrow\infty $, and will be
slightly relevant in the model $WD_{3}^{(p)}$ with $p$ large but
finite. In this case the effect of disorder can reliably studied
by the perturbative renormalization group, by developing in
$\epsilon $.

We observe also that the problem formulated in this way will be
defined better, theoretically, compared to the much used Potts
model, or the minimal conformal theory models, with disorder
[2,3]. In this latter case the disorder is marginal in the case of
the Ising model [4], and the $\epsilon$ parameter of the Potts
model, in which, similarly, the disorder is slightly relevant,
corresponds to the deviation of the central charge, or of the
number of components, from the Ising case. The point of difference
is that in this case there are no unitary models close to Ising
which would correspond to infinitesimal values of $\epsilon$.
Ising model is not the accumulation point of the unitary series.
Next to Ising, the 3-states Potts model, lies at numerically small
but finite distance in $\epsilon$. For this reason, the
perturbative expansion in $\epsilon$ is likely to give numerical
results, in case of Potts model with disorder, not analytic ones.

This is different in case of the model $WD_{n}^{(p)}$ where the
disorder is marginal at the accumulation point of unitary series,
$p=\infty $. In this case for any small value of $\epsilon\sim
1/p$ there would be a physical, unitary model behind. In this
sense, $\epsilon $ will be a well defined analytic parameter. One
could expect that the perturbative expansion in $\epsilon $ would
provide in this case an analytic expansion in $\epsilon = 1/(p+1)$
of unknown yet exact functions (central charge, dimensions of
operators). Like this is actually the case for the minimal model
perturbed with the operator $\phi_{3,1}$ [5,6]. For these reasons
we consider the model to be interesting.

Another important point of difference with the Potts model is that
in the operator product decomposition of
$\varepsilon(x)\varepsilon(x')$ there appears an extra slightly
relevant operator, we shall note it $\phi(x)$. The operator
algebra has the form: \begin{eqnarray}
\varepsilon(x)\varepsilon(x')=\frac{1}{|x-x'|^{2\Delta_{\varepsilon}}}(I+...)\nn\\+
\frac{D_{\varepsilon\varepsilon}^{\phi}}{|x-x'|^{2\Delta_{\varepsilon}-\Delta_{\phi}}}(\phi(x)+...)+
\frac{D_{\varepsilon\varepsilon}^{\phi'}}{|x-x'|^{2\Delta_{\varepsilon}-\Delta_{\phi'}}}(\phi'(x)+...)
\end{eqnarray} The detailed information on this decomposition shall be given
in the sections 3. For the moment we shall observe only that one
of the operators in (1.6), let's say $\phi'(x)$, is irrelevant, so
that it can be dropped in the $RG$ calculation. While another,
$\phi(x)$, is slightly relevant $\Delta_{\phi}=1-A\epsilon $. This
requires, in the $RG$ calculation, to add this extra operator to
the perturbative action, so that the replicated theory will have
the perturbation of the form:
\begin{equation}\label{action}
A_{pert}=g\int d^{2}x\sum^{N}_{a\neq
b}\varepsilon_{a}(x)\varepsilon_{b}(x)+\lambda\int
d^{2}x\sum^{N}_{a=1}\phi_{a}(x)
\end{equation}
with the initial values,in the $RG$ sense, $g=-g_{0},\quad
g_{0}>0$, which is due to disorder, and $\lambda=\lambda_{0}=0$,
initially zero, but produced further spontaneously by the
interactions.

As will be shown in the next Sections, the theory with $A_{pert}$
in (1.7) has a stable fixed point at
\begin{equation}
\lambda_{\ast}\neq 0,\quad g_{\ast}=0
\end{equation}
and it corresponds to $N$ decoupled models $WD_{3}^{(p-1)}$, so
that, in the limit $N=0$ which corresponds to problem with
disorder, one gets a flow:
\begin{equation}
WD_{3}^{(p)}\mbox{with disorder}\rightarrow W_{3}^{(p-1)}
\mbox{regular, without disorder}
\end{equation}
One gets this flow for $N=0$, in the coefficients of the $\beta$ -
function, and for the initial conditions
$g_{0}<0,\quad\lambda_{0}=0.$ There are no other fixed points in
this case $(N=0)$ in addition to (1.8) and the trivial fixed
point, unstable, $g=\lambda=0.$ Yet in the related problem of $N$
coupled regular models (no disorder), for $N=3$ or bigger, the
$\beta$- function contains, in addition to (1.8), two other fixed
points, both stable in one direction and instable in the other.
They correspond to multicriticality. To reach them one has to have
both initial couplings $g_{0},\quad\lambda_{0}\neq 0$, and, in
addition their ratio have to be fine tuned. These extra fixed
points are nontrivial, in the sense of a nontrivial realization at
these points of a permutational symmetry. They are also nontrivial
in the sense that the corresponding couplings
$(g_{\ast}^{(1)},\lambda_{\ast}^{(1)}),\quad (g_{\ast}^{(2)},
\lambda_{\ast}^{(2)})$ are all non-vanishing. We expect that these
points would not be identified with any presently known conformal
field theory, like this appears to be the case for the coupled
Potts models [2,7]. As has been argued above, we expect that in
this case, of coupled $WD_{3}^{(p)}$ and more generally
$WD_{n}^{(p)}$ models, our calculations provide first analytic
corrections in $\epsilon\sim\frac{1}{p}$ to the development of yet
unknown exact functions of $\epsilon $, or $p$. In case of the
Potts models, the corresponding corrections in $\epsilon $ have to
be considered as only numerical.

The rest of the paper is organized as follows.
In Section 2 we introduce the Coulomb-Gas representation of the
$WD_{3}^{(p)}$ model. Then, in Section 3, we study the effects of weak
bond randomness on this model and the related problem of $N$ coupled
$WD_{3}^{(p)}$ models. The renormalization of the couplings and of the
energy operators are computed at one loop order. In Section 4 we study
the  behavior of the RG-flow in the case of disordered system and of
 $N$ coupled systems; we show in particular the presence of non trivial fixed
points in both cases. We give the analytic expansion in
$\epsilon$ of the central charge and of the dimension of energy
operators at these points. In Section 5  all the results we have
found for the $WD_{3}^{(p)}$ model are
generalized for the whole family of $WD_{n}^{(p)}$ models. We discuss
what has been obtained in the Conclusions. 

\section{Coulomb gas of the model $WD_{3}^{(p)}$}

To make our presentation self-contained, and also because the
models $\{ WD_{n}^{(p)}\}$ have not much been used so far in the
statistical physics applications, we shall reproduce in this
Section the results of the paper [1], by giving the details on a
particular model of our interest, $WD_{3}^{(p)}$.

This model could be realized by a 3-component Coulomb gas, with
the stress-energy operator $T(z)$ taking the form:
\begin{equation}
T(z)=-\frac{1}{4}:\partial\vec{\varphi}(z)\partial\vec{\varphi}(z):+i\vec{\alpha}_{0}
\partial^{2}\vec{\varphi}
\end{equation}
Here $\vec{\varphi}(z)=\{\varphi_{1}(z), \varphi_{2}(z),
\varphi_{3}(z)\}$ is a set of three free fields put into a vector,
with the correlation functions normalized as:
\begin{eqnarray}
<\varphi_{a}(z,\bar{z})\varphi_{b}(z',\bar{z'})>=2\log\frac{1}{|z-z'|^{2}}\delta_{ab}\nn\\
=(2\log\frac{1}{z-z'}+2\log\frac{1}{\vec{z}-\vec{z'}})\delta_{ab}
\end{eqnarray}
In the following, and in the above formula (2.1), we often
suppress the dependance of $\varphi_{a}(z,\bar{z}),$ on $\bar{z}$,
which will be implicit.

The vector $\vec{\alpha_{0}}$ in (2.1) corresponds to a presence
of the background charge operator, $V_{-2\vec{\alpha}_{0}}(R)$,
putted at infinity, $R\rightarrow \infty $ [8].

In addition to $T(z)$, the model contains two other operators in
its chiral algebra: $W_{3}(z),\,\,\,W_{4}(z)$, with conformal
dimensions 3 and 4. They could also be expressed in terms of
polynomials in derivatives of fields $\{\varphi_{a}(z)\}$ [1]. The
expressions are longer, but they will not actually be needed in
the present analysis. Important is that $W_{3}(z)$ and $W_{4}(z)$
exist and classify, together with $T(z)$, all the fields
(operators) of the model in terms of primaries and descendants of
the chiral algebra. Then the usual methods, combined with the
available Coulomb gas representation, define dimensions of primary
operators (Kac formula) and their correlation functions. In this
sense the conformal theories of $WD_{n}^{(p)}$ are fully defined
[1].

In the Coulomb gas of the model $WD_{3}^{(p)}$ there are 3
screening operators ``+'', and 3 screening operators ``--'':
\begin{equation} V_{a}^{+}(z)\equiv
V_{\vec{\alpha}_{a}^{+}}(z)=e^{i\vec{\alpha}^{+}_{a}\vec{\varphi}(z)}
\end{equation} \begin{equation} V_{a}^{-}(z)\equiv
V_{\vec{\alpha}_{a}^{-}}(z)=e^{i\vec{\alpha}_{a}^{-}\vec{\varphi}(z)}
\end{equation} \begin{equation}
\vec{\alpha}_{a}^{\pm}=\alpha_{\pm}\vec{e}_{a},\quad
(\vec{e}_{a})^{2}=1 \end{equation} The ``length'' of the screening
operator vectors, $\alpha_{\pm}$, are fixed by the condition that
the conformal dimension of $V_{a}^{\pm}(z)$ is equal to 1. In
general, for a vertex operator \begin{equation}
V_{\vec{\alpha}}(z)=e^{i\vec{\alpha}\vec{\varphi}} \end{equation}
its conformal dimension with respect to the stress-energy operator
$T(z)$, eq.(2.1), is given by: \begin{equation}
\Delta_{\vec{\alpha}}\equiv\Delta(V_{\vec{\alpha}})=\vec{\alpha}^{2}-2\vec{\alpha}\vec{\alpha}_{0}=
\alpha^{2}-2\alpha\alpha_{0}\cos \theta \end{equation} Here
$\Theta$ is the angle between $\vec{\alpha}_{0}$ and
$\vec{\alpha}$. For the geometry of screenings of $WD_{3}^{(p)}$
the angle between $\{\vec{\alpha}_{a}\}$ and $\vec{\alpha}_{0}$ is
the same, Fig.1. One gets, as a condition $\Delta
(V_{a}^{\pm})=1$,
\begin{equation} \alpha^{2}_{a}-2\alpha_{a}\alpha_{0}\cos\Theta=1
\end{equation} \begin{equation}
(\alpha_{a})_{1,2}\equiv\alpha_{\pm}=\alpha_{0}\cos\Theta\pm\sqrt{\alpha_{0}^{2}\cos\Theta+1}
\end{equation}

The unit vectors $\{\vec{e}_{a}\}$ shown in the Fig.1 correspond
to simple roots of the classical Lie algebra $D_{3}$. As it was
said above, the vector $\vec{\alpha}_{0}$ is supposed to be
``equally distant'' from $\vec{e}_{1}, \vec{e}_{2}, \vec{e}_{3}.$
Assuming this condition, for which we will not go into details
here, $\cos\Theta $ will be defined by the geometry in Fig.1. One
finds:
\begin{equation}
\cos\Theta=\frac{1}{\sqrt{10}}
\end{equation}
By the formula (2.9), the lengths $\alpha_{\pm}$ of the screenings
in eq.(2.5) will be functions of $\alpha_{0}$ only. This implies
that in the model $WD_{3}^{(p)}$ the orientational geometry is
fixed, but there is one free parameter in the lengths of the
vectors: the screenings $\{\vec\alpha_{a}^{(\pm)}\}$ and the
background charge $\vec\alpha_{0}.$

The primary operators of the model are represented by the vertex
operators \begin{equation}
V_{\vec{\beta}}(z)=e^{i\vec{\beta}\vec{\varphi}(z)} \end{equation}
The allowed values of the vectors $\vec{\beta}$ are defined by the
degeneracy condition of the modules of $V_{\vec{\beta}}(z)$ with
respect to the chiral algebra. They are found to be given by [1]:
\begin{equation}
\vec{\beta}=\vec{\beta}_{(n'_{1},n_{1})(n'_{2},n_{2})(n'_{3},n_{3})}=\sum_{a=1}^{3}(\frac{1-n'_{a}}{2}
\alpha_{-}+\frac{1-n_{a}}{2}\alpha_{+})\vec{\omega}_{a}
\end{equation} Here $\{\vec{\omega}_{a}\}$ is a set of three
vectors which are dual to the unit vectors of the screenings
$\{\vec{e}_{a}\}$:
\begin{equation} (\vec{\omega}_{a},\vec{e}_{b})=\delta_{ab} \end{equation}
Using the general formula for the dimensions of vertex operators,
eq.(2.7), one then gets from (2.12) the Kac formula of the model
which defines the set of dimensions of the primary operators
$\{\phi_{(n'_{1},n_{1})(n'_{2},n_{2})(n'_{3},n_{3})}(z)\}:$
\begin{equation}
\phi_{(n'_{1},n_{1})(n'_{2},n_{2})(n'_{3},n_{3})}(z)\propto
V_{\vec{\beta}_{(n'_{1},n_{1})(n'_{2},n_{2}) (n'_{3},n_{3})}}(z)
\end{equation}
\begin{eqnarray}
\Delta_{(n'_{1},n_{1})(n'_{2},n_{2})(n'_{3},n_{3})}=\vec{\beta}_{(...)}^{2}-2\vec{\alpha}_{0}\vec
{\beta}_{(...)}\nn\\=\sum_{a}(u(n'_{a},n_{a}))^{2}\frac{(\vec{\omega}_{a})^{2}}{4}+2\sum_{a<b}u(n'_{a},
n_{a})u(n'_{b},n_{b})\frac{(\vec{\omega}_{a},\vec{\omega}_{b})}{4}\nn\\-(\alpha_{+}+\alpha_{-})
\sum_{a,b}u(n'_{a},n_{a})\frac{(\vec{\omega}_{a},\vec{\omega}_{b})}{2}
\end{eqnarray} Here
\begin{equation}
u(n'_{a},n_{a})=(1-n'_{a})\alpha_{-}+(1-n_{a})\alpha_{+}
\end{equation} In (15) we have used in addition the decomposition:
\begin{equation}
2\vec{\alpha}_{0}=(\alpha_{+}+\alpha_{-})\sum_{a}\vec{\omega}_{a}
\end{equation} which is easily verified by multiplying both sides with a
vector $\vec{e}_{b}$ and using: \begin{equation}
2\vec{\alpha}_{0}\vec{e}_{b}=2\alpha_{0}\cos\Theta=\alpha_{+}+\alpha_{-}
\end{equation} -- by eq.(2.9), and
$(\vec{\omega}_{a}\vec{e}_{b})=\delta_{ab}$, eq.(2.13). The unit
vectors $\{\vec{e}_{a}\}$, which are defined by the Fig.1, when
expressed in components will be given by: \begin{equation}
\vec{e}_{1}=\frac{1}{\sqrt{2}}(0,-1,1),\quad
\vec{e}_{2}=\frac{1}{\sqrt{2}}(1,1,0),\quad \vec{e}_{3}=
\frac{1}{\sqrt{2}}(-1,1,2) \end{equation} The dual vectors
$\{\vec{\omega}_{a}\}$, defined by the eq.(2.13), are found to be
equal to:
\begin{equation}
\vec{\omega}_{1}=\sqrt{2}(0,0,1),\quad
\vec{\omega}_{2}=\frac{1}{\sqrt{2}}(1,1,1),\quad \vec{\omega}_{3}=
\frac{1}{\sqrt{2}}(-1,1,1)
\end{equation} Then one finds:
\begin{equation}
(\vec{\omega}_{a},\vec{\omega}_{b})=\left(\begin{array}{ccc}2&1&1\\1&\frac{3}{2}&\frac{1}{2}\\
1&\frac{1}{2}&\frac{3}{2}\end{array}\right)
\end{equation}
Finally, the Kac formula of the model $WD_{3}^{(p)}$ takes the
following form:
\begin{eqnarray}
\Delta_{(n'_{1},n_{1})(n'_{2},n_{2})(n'_{3},n_{3})}=\frac{1}{2}(u(n'_{1},n_{1}))^{2}+\frac{3}{8}
(u(n'_{2},n'_{2}))^{2}+\frac{3}{8}(u(n'_{3},n'_{3}))^{2}\nn\\+\frac{1}{2}u(n'_{1},n_{1})u(n'_{2},
n_{2})+\frac{1}{2}u(n'_{1},n_{1})u(n'_{3},n_{3})+\frac{1}{4}u(n'_{2},n_{2})u(n'_{3},n_{3})\nn\\-
(\alpha_{+}+\alpha_{-})(2u(n'_{1},n_{1})+\frac{3}{2}u(n'_{2},n_{2})+\frac{3}{2}u(n'_{3},n_{3}))
\end{eqnarray} $u(n'_{a},n_{a})$ are defined by eq.(2.16). One checks that
\begin{equation} \sum_{b}(\vec{\omega}_{a},\vec{\omega}_{b})=(4,3,3) \end{equation}
which we have used in the last term in (2.22).

As it was mentioned above, the model contains one free parameter.
This will be either $\alpha_{0}$, or $\alpha_{+}$, or
$\alpha_{-}$, or the central charge of the model, which, for
$T(z)$ in eq.(2.1), will be given by:
\begin{equation}
c=3-24\vec{\alpha}_{0}^{2}
\end{equation}
One observes that the lengths of the screenings ``+'' and the
screenings ``--'', $\alpha_{+}$ and $\alpha_{-}$ in eq.(2.9), are
related:
\begin{equation}
\alpha_{+}+\alpha_{-}=2\alpha_{0}\cos\Theta
\end{equation}
\begin{equation}
\alpha_{+}\alpha_{-}=-1
\end{equation}
Similar to the case of the basic conformal theory of [9], one gets
a unitary set of models if the parameter $(\alpha_{+})^{2}$ is
restricted to the discrete values:
\begin{equation}
(\alpha_{+})^{2}=\frac{p+1}{p}
\end{equation}
By the relation (2.26) one will have also
\begin{equation}
(\alpha_{-})^{2}=\frac{p}{p+1}
\end{equation}
For $WD_{3}^{(p)},\,\,\,p=5,6,...$ [1]. In the following we shall
restrict ourselves to this unitary series. This explains for the
extra index $p$ of $WD_{3}^{(p)}$.

We shall next be interested in this model in the limit of $p$
large, $p\gg 1$. First, when we take $p\rightarrow\infty $, the
Kac formula (2.22) simplifies and takes the form:
\begin{equation}
\alpha_{+}=1,\quad\alpha=-1
\end{equation}
\begin{equation}
u(n'_{a},n_{a})=n'_{a}-n_{a}
\end{equation}
\begin{eqnarray}
\Delta_{(n'_{1},n_{1})(n'_{2},n_{2})(n'_{3},n_{3})}=\frac{1}{2}(n'_{1}-
n_{1})^{2}+\frac{3}{8}(n'_{2}-n_{2})^{2}+\frac{3}{8}(n'_{3}-n_{3})^{2}\nn\\+
\frac{1}{2}(n'_{1}-n_{1})(n'_{2}-n_{2})+\frac{1}{2}(n'_{1}-n_{1})(n'_{3}-
n_{3})+\frac{1}{4}(n'_{2}-n_{2})(n'_{3}-n_{3})
\end{eqnarray}
In general, not specifically for $p$ large or infinite, the model
contains a symmetry, in the set of its primary operators and their
dimensions, w.r.t. permutation of indexes 2 and 3. This
corresponds to the reflectional $Z_{2}$ symmetry of the geometry of
the screening vectors $\{\vec{e}_{a}\}$ in Fig.1.

In the set of dimensions
$\Delta_{(n'_{1},n_{1})(n'_{2},n_{2})(n'_{3},n_{3})}$ there is a
subset which is not $Z_{2}$ degenerate and which corresponds to
the singlet operators. For instance, singlet operators are
$\{\phi_{(n'_{1}, n_{1})(1,1)(1,1)}\}$. The lowest dimension
operator in the singlet sector, next to the identity operator
$I=\phi_{(1,1)(1,1)(1,1)}(\Delta_{(1,1)(1,1)(1,1)}=0,$ by
eq.(2.22)), will be the operator \begin{equation}
\phi_{(2,1)(1,1)(1,1)}(x) \end{equation} It is natural to identify
it with the energy operator of the model,
\begin{equation}
\varepsilon(x)=\phi_{(2,1)(1,1)(1,1)}(x)
\end{equation}
If we look now at the limiting case of $p\rightarrow\infty $, we
shall find, by the formula (2.31), that
\begin{equation}
\Delta_{\varepsilon}(p\rightarrow\infty)=\frac{1}{2}
\end{equation}
For $p$ finite, by the formula (2.22), one finds:
\begin{equation}
\Delta_{\varepsilon}=\Delta_{(2,1)(1,1)(1,1)}=\frac{5}{2}\alpha_{-}^{2}-2
\end{equation} Substituting, by eq.(2.28),
\begin{equation}
\alpha_{-}^{2}=\frac{p}{p+1}=1-\frac{1}{p+1}=1-\epsilon
\end{equation}
where we have defined
\begin{equation}
\epsilon=\frac{1}{p+1}
\end{equation} one obtains:
\begin{equation}
\Delta_{\varepsilon}=\frac{1}{2}-\frac{5}{2}\epsilon
\end{equation}
In our present study we are going to couple disorder to the energy
operator $\varepsilon(x)$, defined in eq.(2.33). In the related
problem of coupled regular models, we shall couple the models
between themselves by their energy operators:
\begin{equation}
A_{pert}=g\sum_{a\neq
b}^{N}\int d^{2}x\:\varepsilon_{a}(x)\varepsilon_{b}(x)
\end{equation}
As has been discussed in the introduction, in both cases one gets
a problem with a slightly relevant perturbation, if
$\Delta_{\varepsilon}$ is given by the eq.(2.38) with $\epsilon$
small, or $p$ large, eq.(2.37). These problems will next be
studied by the methods of the perturbative RG.

\newpage

\section{1-loop RG equations for $WD_{3}^{(p)}$}
\subsection{Renormalization of the couplings}
We initially consider $N$ regular $WD_{3}^{(p)}$ models coupled by the
energy-energy interaction (2.39):
\begin{equation}
A=\sum_{a}^{N} A_{0}^{(a)} + g_{0}\sum_{a\neq b}^{N}\int d^{2}x
\: \varepsilon_{a}(x)\varepsilon_{b}(x)
\end{equation}
$A_{0}^{(a)}$ being the conformal action corresponding to a single
$WD_{3}^{(p)}$ model. The action (3.1) describes a conformal field
theory perturbed by a slightly relevant term quadratic in the
energy operator; such a problem can be reliably studied by means
of perturbative RG with $\epsilon$-expansion.  The RG scheme
requires that all the relevant terms produced by the energy-energy
interaction have to be added to (3.1) and  the algebra of
the enlarged set of perturbing fields have not to present other
relevant operators. As shown
below, the O.P.E. of $\varepsilon(x)\varepsilon(x')$ contains,
apart from identity, one and only one (slightly) relevant
operator, namely $\phi_{(1,1)(2,1)(2,1)}\equiv\phi$, whose
dimension is
 $\Delta_{(1,1)(2,1)(2,1)}=1-4\epsilon$.  This implies  that in the perturbative
computation there will be diagrams (generated by terms in (2.39) with
the same replica index) which produce, apart from trivial or irrelevant
contributions, the term:
\begin{equation}
\sum_{a}^{N}\int d^{2}x\: \phi_{a}(x)
\end{equation}
We therefore consider the more general
action:
\beq
A=\sum_{a}^{N} A_{0}^{(a)} + g_{0}\sum_{a\neq b}^{N}\int d^{2}x
\: \varepsilon_{a}(x)\varepsilon_{b}(x)+\lambda_{0}\sum_{a}^{N}\int d^{2}x\: \phi_{a}(x)
\eeq
The problem of a
single $WD_{3}^{(p)}$ model perturbed by $\phi$ has been considered in
[1], where the existence of a non trivial infrared fixed
point to which the system flows has been shown. The conformal field theory associated
to this point corresponds to the $WD_{3}^{(p-1)}$ model.

In order to  investigate the energy algebra, we study the
  four point correlation function $G(x)\equiv\langle \varepsilon(0) \varepsilon(x) \varepsilon(1)
\varepsilon(\infty) \rangle$, which can be decomposed in a sum over
  s-channel diagrams corresponding to insertions of different
  operators.
Introducing the Coulomb gas representation (\textit{cfr.} section
2) for  $G(x)$ and  defining for simplicity
$\vec{\beta}_{(2,1)(1,1)(1,1)} \equiv \vec{\beta}_{\varepsilon}$,
$2\vec \alpha_{0}-\vec \beta_{(2,1)(1,1)(1,1)}\equiv
\vec{\beta}_{\bar{\varepsilon}}$,
$\vec{\beta}_{(1,1)(2,1)(2,1)}\equiv \vec{\beta}_{\phi}$, and
$2\vec \alpha_{0}-\vec \beta_{(1,1)(2,1)(2,1)}\equiv
\vec{\beta}_{\bar{\phi}}$.  we can write : \beq \label{integral}
 G(x)\propto\int..\int \langle
V_{\vec{\beta_{\varepsilon}}}(0)V_{\vec{\beta_{\varepsilon}}}(x)V_{\vec{\beta_{\varepsilon}}}(1)
V_{\vec{\beta_{\bar{\varepsilon}}}}(\infty)V_{1}^{-}(\mu_{1})V_{1}^{-}(\mu_{2})V_{2}^{-}(\xi)V_{3}^{-}(\nu)\rangle
d^{2}\mu_{1} d^{2}\mu_{2} d^{2}\xi d^{2}\nu
\eeq where the screenings
$V_{a}^{-}$, defined in (2.3), are integrated over the $2D$ plane and
their number is determined by the  charge neutrality condition :
\beq
2\vec{{\beta}_{\varepsilon}}+\alpha_{-}\left(m_{1}\vec{e}_{1}+m_{2}\vec{e}_{2}+m_{3}\vec{e}_{3}\right)=0
\label{neutralitycharge} \eeq
which is satisfied for $m_{1}=2$, $m_{2}=1$,
$m_{3}=1$.

 Each leading term in $x\to 0$ limit  originates in
the integral representation (3.4) when a particular number $q_{a}$
of screenings $V_{a}^{-}$ (by eq.(3.5) $q_{1}\leq 2$, $q_{2}\leq 1$,
$q_{3}\leq 1$) approaches the origin. Since in this limit \beq
V_{\vec{\beta}_{\varepsilon}}(x)
V_{\vec{\beta}_{\varepsilon}}(0)\prod_{a}(V_{a}^{-})^{q_{a}}\to
V_{2\vec{\beta}_{\varepsilon}+\alpha_{-}\sum_{a}
q_{a}\vec{e}_{a}}(0)+\cdots, \eeq the intermediate state vertex
$V_{2\vec{\beta}_{\varepsilon}+\alpha_{-}\sum_{a}
q_{a}\vec{e}_{a}}$ corresponds to a primary field which belongs to
the energy algebra.
 In particular we have noticed the presence of the vertex $V_{2\vec{\beta}_{\varepsilon}
+\alpha_{-}\vec{e}_{1}}$, which corresponds to the primary operator
$\phi$, according to
\beq
2\vec{\beta}_{\varepsilon}
+\alpha_{-}\vec{e}_{1}=-\alpha_{-}\left(\frac{\vec{\omega}_{2}}{2}+\frac{\vec{\omega}_{3}}{2}\right)
\eeq
In appendix A we shall show in detail that $\phi$ is the only
relevant operator in the energy product decomposition and that the
 algebra of $\phi$ and $\varepsilon$, apart from irrelevant terms, is:
\begin{eqnarray}
\varepsilon(x)\varepsilon(0)&=&\frac{I}{|x|^{4\Delta_{\varepsilon}}}
+\frac{D^{\phi}_{\varepsilon\varepsilon}}{|x|^{4\Delta_{\varepsilon}-2\Delta_{\phi}}}\phi+\cdots \nonumber \\
\phi(x)\phi(0)&=&\frac{I}{|x|^{4\Delta_{\phi}}}+\frac{D^{\phi}_{\phi\phi}}{|x|^{2\Delta_{\phi}}}\phi+\cdots\nonumber\\
\phi(x)\varepsilon(0)&=&\frac{D^{\varepsilon}_{\varepsilon\phi}}{|x|^{2\Delta_{\phi}}}\phi+\cdots;
\end{eqnarray}
$D^{\phi}_{\varepsilon\varepsilon}$ and $D^{\phi}_{\phi\phi}$ are the
associated structure constants of the $WD_{3}^{(p)}$ model.

At one loop order the main RG characteristics of action (3.3), e.g.
$\beta$-functions or  renormalization of operators, are easily
obtained from eq.(3.8) (see appendix C). The renormalized coupling constants $g(r)$
and $\lambda(r)$ ($r$ is the short distance cut-off) have been
computed, and the correspondent $\beta$-functions take the form:
\begin{eqnarray}
         \beta_{g}\equiv \frac{\partial g(r)}{\partial
         ln(r)}&=&(2-4\Delta_{\varepsilon})
         g-(N-2)g^{2}-4D_{\varepsilon\varepsilon}^{\phi}\lambda g \nonumber\\
         \beta_{\lambda}\equiv \frac{\partial\lambda(r)}{\partial
ln(r)}&=&
        (2-2\Delta_{\phi})\lambda-\frac{D_{\phi\phi}^{\phi}}{2}
       \lambda^{2}-\frac{N-1}{2}D_{\varepsilon\varepsilon}^{\phi}g^{2}
\end{eqnarray}
where  the coupling constants have been redefined as $g\to g/4\pi$
and $\lambda \to\lambda/ 2\pi$. In order to study the coupling
flow induced by eq.(3.9) (section 4) the structure constants have been
determined (see appendix B): \beq
D^{\phi}_{\varepsilon\varepsilon}=\sqrt{\frac{5}{3}}+0(\epsilon),\quad
D^{\phi}_{\phi\phi}=\frac{8}{\sqrt{15}}+0(\epsilon) \eeq

In section 4 we will investigate the properties of equations
(3.9)  for $N=0$ (disorder) and $N\geq 2$ (coupled systems).

\subsection{Renormalization of the energy operator}
In order to  study the effect of the perturbation on the energy
operators, we need to compute the renormalized operators
$\varepsilon_{a}'$, which are expressed via the $N$x$N$ matrix
$[Z_{\varepsilon}]_{ab}$ by: \beq
\varepsilon_{a}'=\sum_{b}[Z_{\varepsilon}]_{ab}\:\varepsilon_{b}
\eeq As usual, we proceed perturbatively: computing contributions
from each coupling term and rewriting bare quantities in term of
 renormalized ones, we
 have determined  up to the first order
 the matrix $\gamma_{ab}\equiv d
 ln[Z]_{ab}/d
ln(r)$ (see appendix C). The new critical exponents at the fixed
   points $g_{*},\lambda_{*}$ of the RG flow will be given by the eigenvalues of  the
dimension matrix $\Delta_{ab}\equiv
\Delta_{\varepsilon }\delta_{ab}-\gamma_{ab}(g_{*},\lambda_{*})$ which takes the form:
\beq
\label{dimmatrix}
\Delta_{ab}=\left[\matrix{\Delta_{\varepsilon_{1}}+D^{\phi}_{\varepsilon\phi}\lambda_{*}
   & g_{*}& g_{*}&\cdot&\cdot& g_{*}
   \cr   g_{*} &\Delta_{\varepsilon_{2}}+D^{\phi}_{\varepsilon\phi}\lambda_{*}&g_{*}&\cdot&\cdot& g_{*}
   \cr g_{*} & g_{*}&\Delta_{\varepsilon_{3}}+D^{\phi}_{\varepsilon\phi}\lambda_{*}& g_{*}&\cdot&g_{*}
 \cr \cdot& \cdot &g_{*}&\cdot& \cdot&\cdot&
 \cr \cdot& \cdot & \cdot &\cdot& \cdot&\cdot&
 \cr g_{*}& g_{*}&\cdot&\cdot&\cdot&
  \Delta_{\varepsilon_{N}} +D^{\phi}_{\varepsilon\phi}\lambda_{*}}\right]
\eeq
It is straightforward to see that the symmetric combination
$\varepsilon_{s}=\varepsilon_{1}+\varepsilon_{2}+\ldots+\varepsilon_{N}$
and the antisymmetric ones
$\varepsilon_{a_{1}}=\varepsilon_{1}-\varepsilon_{2}$ ,
$\varepsilon_{a_{2}}=\varepsilon_{2}-\varepsilon_{3}\ldots$
are eigenvectors of (3.12) and
 their dimensions turn out to be:
\begin{eqnarray}
\Delta_{\varepsilon_{s}}(g_{*},\lambda_{*})&=&\Delta_{\varepsilon}+D^{\varepsilon}_{\varepsilon\phi}\lambda_{*}+(N-1)g_{*}
\nonumber \\
\Delta_{\varepsilon_{a_{1}}}(g_{*},\lambda_{*})&=&\Delta_{\varepsilon_{a_{2}}}=\ldots=\Delta_{\varepsilon}+D^{\varepsilon}_{\varepsilon\phi}\lambda_{*}-g_{*}
\end{eqnarray}
In the next section we shall show that equations (3.9) admit
fixed points with $g_{*}\neq 0$ at which $N$ identical systems remain
coupled; we expect therefore that the correspondent
conformal field theory realizes in a non trivial way the permutation
symmetry of $N$ identical objects. The splitting of
dimensions in the energy sector $\Delta_{\varepsilon_{s}}\neq \Delta_{\varepsilon_{a}}$
indicates that this is indeed the case. Although a similar result was obtained for $N$
coupled Potts models, in this case we have access to the analytic expansion
in $\epsilon$ of the dimensions of the energy operators and of the
central charge (see next section) of new
conformal field theories.
\section{RG-flow and central charge}
\subsection{Fixed point structure and couplings flow}
The first step in the study of RG-flow is to find all points $
g_{*},\lambda_{*}$  such that  $\beta_{g}(g_{*},\lambda_{*})=\beta_{\lambda}(g_{*},\lambda_{*})=0$.\\
The equations (3.9) have been obtained for a generic number $N$ of
coupled models. We consider firstly 
the quenched case which is obtained in the $N\to 0$ limit.
The $\beta$-functions, defined in (3.9), take in this limit the form: 
\begin{eqnarray}
\beta_{g}&=&10\epsilon
g+2g^{2}-2\sqrt{\frac{5}{3}}\lambda g \nonumber \\
 \beta_{\lambda}&=&8\epsilon\lambda-\frac{4}{\sqrt{15}}
\lambda^{2}+\frac{1}{2}\sqrt{\frac{5}{3}}g^{2}
\end{eqnarray}
 The RG flow has, apart from the trivial solution $g=\lambda=0$,
one stationary point:
\beq\label{origin}
\lambda_{*}=2\sqrt{15}\epsilon,\quad g_{*}=0 \label{zam} \eeq
which is stable in all directions. 
This point has already been found in [1]: the associated  critical
theory is described by a pure $WD_{3}^{(p-1)}$ model. Taking  as initial conditions
 $\lambda_{0}=0$ and $g_{0}< 0$, and
 assuming that higher loop corrections will not change the
qualitative behavior of the flow near the two fixed points, the
system 
 flows toward point (\ref{zam}) as supported by the numerical
 calculations shown in Fig. 2.\\
 Initial conditions $\lambda_{0}=0$ and $g_{0}< 0$ correspond 
exactly to the disorder produced couplings; so, the addition of a
small bond randomness will drive the  model $WD_{3}^{(p)}$ to the model
 $WD_{3}^{(p-1)}$ without disorder.

We consider now  $N\geq 2$, i.e. the case of coupled models: by (3.9) and (3.10) the
$\beta$-functions are:
\begin{eqnarray}
\beta_{g}&=&10\epsilon g-(N-2)g^{2}-2\sqrt{\frac{5}{3}}\lambda g
\nonumber \\
\beta_{\lambda}&=&8\epsilon\lambda-\frac{4}{\sqrt{15}}
\lambda^{2}-\frac{N-1}{2}\sqrt{\frac{5}{3}}g^{2} 
\end{eqnarray}
In addition to point (4.2), which remains a point of
attraction of the RG-flow  since the respective RG-eigenvalues
don't depend on $N$, we have two new fixed points:
\beq
\lambda_{*}^{(1)}=\sqrt{15}\left(1-(N-2)\sqrt{\frac{6}{N-1+6N^{2}}}\right)\epsilon,\quad
g_{*}^{(1)}=10\sqrt{\frac{6}{N-1+6N^{2}}}\epsilon \label{B} \eeq \beq
\lambda_{*}^{(2)}=\sqrt{15}\left(1+(N-2)\sqrt{\frac{6}{N-1+6N^{2}}}\right)\epsilon,\quad
g_{*}^{(2)}=-10\sqrt{\frac{6}{N-1+6N^{2}}}\epsilon \label{C} \eeq
The stationary points (\ref{B}) and (\ref{C}) constitute
node points of the flow  and they can be reached only by a
fine tuning of the initial conditions. Studying in detail the flow
diagram of (4.3) (shown for $N=3$ in Fig.3)  we see that with the  initial conditions
 $\lambda_{0}=0$ and $g_{0}\neq 0$ the system  will flow far from our fixed points
toward either a massive theory or another fixed point which cannot
be seen at this order in perturbation theory\footnote{It could be mentioned that a particular case of
$N=2$ coupled $WD^{(p)}_{n}$ theories, with
$g_{0}\neq 0, \lambda_{0}=0$, have been considered
previously in the paper [11], with a conclusion that
the theory is integrable. We only want to mention
again that according to our RG analysis  the
second coupling, $\lambda\neq 0$,  will have to be
admitted eventually into the action (see the
corresponding flows in Fig.3).
 
This is in contrast with two coupled Virasoro
algebra minimal models $M_{p}$ which were  shown to be
integrable [12] and studied in detail in [11].
In this last case the original action, with a
coupling term g$_{0}$ only, is stable with respect to RG evolution.
 
Then, if the coupling $\lambda\neq 0$ has to be
taken into the action, for the two coupled
$WD^{(p)}_{n}$ theories, eq.(3.3), the natural
question would be if the presence
of the second perturbative term will modify essentially the analysis of integrability.}.
On the other hand, in the region  $\lambda >0$ and of order $\epsilon$, the
situation is rather different. There are two solutions of eq.(4.3), say
$g_{+}(\lambda)>0$ and $g_{-}(\lambda)<0$, which are attracted
respectively by the node point (4.4) and (4.5); if we take  $\lambda_{0}>0$ and
$g_{-}(\lambda_{0})<g_{0}<g_{+}(\lambda_{0})$, the system will flow
toward the stable point (4.2). In this case the infrared limit of the
system will be described by $N$ decoupled $WD_{3}^{(p-1)}$ models.
     
It's important to note that, although quite similar, the flow is not symmetric along the
$\lambda$ axis, and this will be at the origin of  the difference
between the values of the central charge at the two fixed
points. This asymmetry can be explained by a simple physical argument.
Indeed, when $g>0$, the coupling of different models is always frustrated, while
for $g<0$ the $N$-system can arrange itself in a kind of ferromagnetic
configuration in order to minimize the energy. For $N=2$ (i.e no frustration) we recover the
symmetry $g\rightarrow -g$, and the central charge has the same
value at the two new fixed points (see next section).

Inserting the values of the structure constants (3.10) in eq.
(3.13), we have at the fixed points $(g_{*}^{(1)},\lambda_{*}^{(1)})$
 and $(g_{*}^{(2)},\lambda_{*}^{(2)})$: 
\begin{eqnarray}
\Delta_{\varepsilon_{s}}^{(1),(2)}&=&\Delta_{\varepsilon}+5\left(1\pm
N\sqrt{\frac{6}{N-1+6N^{2}}}\right)\epsilon\nonumber \\
\Delta_{\varepsilon_{a}}^{(1),(2)}&=&\Delta_{\varepsilon}+5\left(1\mp
N\sqrt{\frac{6}{N-1+6N^{2}}}\right)\epsilon
\end{eqnarray}
where
$\Delta_{\varepsilon_{s,a}}^{(1)}\equiv\Delta_{\varepsilon_{s,a}}(g_{*}^{(1)},\lambda_{*}^{(1)})$ 
 and
$\Delta_{\varepsilon_{s,a}}^{(2)}\equiv\Delta_{\varepsilon_{s,a}}(g_{*}^{(2)},\lambda_{*}^{(2)})$.
By eq. (4.6) we have $
\Delta_{\varepsilon_{s}}^{(1)}=\Delta_{\varepsilon_{a}}^{(2)}$ and 
$\Delta_{\varepsilon_{a}}^{(1)}=\Delta_{\varepsilon_{s}}^{(2)}$ for
all $N$. This result can be easily explained for $N=2$: in
this case the two models are equivalent under the replacements  $g\to-g$ and $\varepsilon_{1}\to-\varepsilon_{1}$. On the other hand,
for $N>2$ no reason can be given, as it can be seen
from the multiplicity of the antisymmetric energy combinations. We
believe therefore that these equalities are accidental and they originate
from the simplicity of the 
one-loop order computation.

\subsection{Central charge}
In the previous section we have shown that the RG-flow of the
quenched system ($N=0$) and of the coupled systems ($N\geq 2$) exhibits
non-trivial fixed points. A simple way for computing the central
charge of the associated conformal theories is given by the
Zamolodchikov's c-theorem \cite{zamo}; the theorem provides us with a function
of the couplings $c(g,\lambda)$, to be defined below, whose value
$c(g_{*},\lambda_{*})$ at the fixed point $(g_{*},\lambda_{*})$ is
equal to 
the central charge of the corresponding critical theory.

We define as $\Theta(x)$ the trace of the stress-energy tensor. It's
well known that $\Theta(x)$, which is zero at the fixed point, is
proportional to the perturbing terms of the theory. In our
case we 
have from the 
action (3.3):
\beq
\Theta(x)= \frac{\beta_{g}}{8}\sum_{a\neq
b}^{N}\varepsilon_{a}(x)\varepsilon_{b}(x)+\frac{\beta_{\lambda}}{4}\sum_{c}^{N}\phi_{c}(x)
\eeq
where the renormalization  of the couplings $g\to g/4\pi$
and $\lambda \to\lambda/ 2\pi$ is taken into account.  The function
$c(g,\lambda)$ is then completely determined by the following
equations \cite{zamo}:
\begin{eqnarray}
&&\beta_{g}\frac{\partial c(g,\lambda)}{\partial
g}+\beta_{\lambda}\frac{\partial c(g,\lambda)}{\partial
\lambda}=-24<\Theta(0)\Theta(1)>\nonumber\\
&&c(0,0)=c_{pure}
\end{eqnarray}
with $c_{pure}$ and $<\Theta(0)\Theta(1)>$ respectively the central
charge and the $\Theta$ field two-point correlation function of the
unperturbed theory. 
Inserting eq.(4.7) and eq.(3.9) into eq.(4.8) and using the following
relations:
\begin{eqnarray}
&&\sum^{N}_{a\neq b}\sum^{N}_{c\neq
d}<\varepsilon_{a}(1)\varepsilon_{b}(1)
\varepsilon_{c}(0)\varepsilon_{d}(0)>=2N(N-1)\nonumber \\
&&\sum^{N}_{c\neq
d}\sum^{N}_{a=1}<\varepsilon_{c}(0)\varepsilon_{d}(0)\phi_{a}(1)>=0
\nonumber \\
&&\sum^{N}_{a=1}\sum^{N}_{b=1}<\phi_{a}(1)\phi_{b}(0)>=N,
\label{metric}
\end{eqnarray}
it's straightforward to verify that the function
\beq
c(g,\lambda)=c_{pure}-N\left(\frac{3}{2}\Delta_{\varepsilon}\epsilon g^{2}
+3\Delta_{\phi}\epsilon\lambda^{2}-\frac{(N-1)(N-2)}{4}g^{3}-\frac{D_{\phi \phi}^{\phi}}{4}\lambda^{3}
-\frac{3}{4}(N-1)D_{\varepsilon \varepsilon}^{\phi}g^{2}\lambda
\right) \label{cc}
\eeq
satisfies eq.(4.8). The function (4.10) has been obtained when a number $N$ of
$WD_{3}^{(p)}$ models is considered and so
$c_{pure}=Nc(WD_{3}^{(p)})$, where $c(WD_{3}^{(p)})$ is the central
charge of a single $WD_{3}^{(p)}$ model, calculated in [1]:
\beq
c(WD_{3}^{(p)})=3\left(1-\frac{20}{p(p+1)}\right) \label{charge}
\eeq
In order to compute the central charge in the case of the
disordered problem, 
we normalize the function $c(g,\lambda)$ by
$N$ and then we take the limit $N\to 0$; in fact, in terms of replicas this is
 exactly the limit in which the related quenched free energy is
obtained.
Using the values of the structure constants (3.10), the central charge $c_{dis.}$ at the infrared stable point (4.2) turns
out to be:
\beq
c_{dis.}=\lim_{N\to 0} \frac{c(0,2\sqrt{15}\epsilon)}{N}=c(WD_{3}^{(p)})-120\epsilon^{3}\approx
 c(WD_{3}^{(p-1)})
\eeq
This result is consistent with  what has already been said in the previous section:
the infra-red behavior of a $WD_{3}^{(p)}$ model with disorder
coupled to the energy operator is described by a $WD_{3}^{(p-1)}$
model.

In the related problem of $N$ coupled models, the central charge
$c_{coupl.}$ at
the fixed point (4.2) is
$c_{coupl.}=c(0,2\sqrt{15}\epsilon)=Nc(WD_{3}^{(p-1)})$, i.e.  the
correspondent critical theory is described by $N$ decoupled
$WD_{3}^{(p-1)}$ models. 

Finally we have  access to the analytic
expansion up to the third order in $\epsilon$ of
the central charges $c_{1}$ and $c_{2}$ at the new fixed points
$(g_{*}^{(1)},\lambda_{*}^{(1)})$ and  $(g_{*}^{(2)},\lambda_{*}^{(2)})$:
\beq
c_{1,2}\simeq Nc(WD_{3}^{(p)})-N\left[60\left(1\mp(N-2)\sqrt{\frac{6}{N-1+6N^{2}}}\right)\right]\epsilon^{3}
\label{cbc} \eeq
The eq.(4.13) represents the first analytic result for a new series of
conformal theories which in addition to the $W$-symmetry  present
a non-trivial representation of the permutation symmetry.  

\section{Analysis of the $WD_{n}^{(p)}$ models}
We shall show that the results we have obtained in the previous
sections still hold in the case of $WD_{n}^{(p)}$ models. The
construction of the  Coulomb-Gas representation of these theories
is a direct generalization of what
 has been  presented in Section 2. The  $WD_{n}^{(p)}$ model
presents, in addition to the conformal symmetry, a series of
additional symmetries which are generated by a series of local
currents \{$W_{2k}(x)$\}$_{k=1,\ldots,n-1}$ and $W_{n}(x)$ with 
 dimension $\Delta_{2k}=2k$ and $\Delta_{n}=n$ respectively. It can be
represented by an 
$n$-component Coulomb Gas. The (2.3), (2.4) and (2.5)  still
define the screenings ``+'' and ``$-$'',
 with unit vectors $\vec{e}_{a}$ which lie on a $n$-dimensional space
and correspond to simple roots of Lie algebra $D_{n}$. The primary
operators  are represented by vertex operators $V_{\vec{\beta}}$
with \beq
\vec{\beta}\equiv\vec{\beta}_{(n'_{1},n_{1})(n'_{2},n_{2})..(n'_{n},n_{n})}=\sum_{a=1}^{n}(\frac{1-n'_{a}}{2}
\alpha_{-}+\frac{1-n_{a}}{2}\alpha_{+})\vec{\omega}_{a} \eeq where 
the set of dual vectors $\{\vec{\omega}_{a}\}$, defined
by (2.13),
 have  the quadratic form matrix $F_{ab}\equiv(\vec{\omega}_{a},\vec{\omega}_{b})$:
\begin{eqnarray}
F_{ab}&=& 2a,\quad a\leq b<n-1;\quad F_{a n-1}=F_{a n}=a,\quad a<n-1; \nonumber \\
F_{nn}&=&F_{n-1 n-1}=\frac{n}{2};\quad F_{n-1 m}=\frac{n-2}{2}
\label{quadratic}
\end{eqnarray}
The central charge $c(WD_{n}^{(p)})$ of the $WD_{n}^{(p)}$ model
 has been calculated in [1]: \beq
c(WD_{n}^{(p)})=n\left(1-\frac{(2n-2)(2n-1)}{p(p+1)}\right)
\label{cn} \eeq Using (2.15) and (\ref{quadratic}) the dimension
of primary operators can be easily calculated. In particular the
dimension 
of the operator
$\Phi_{(2,1),(1,1),..,(1,1)}$, naturally identified with the
energy operator $\varepsilon$ (Section 2), is: \beq
\Delta_{\varepsilon}\equiv\Delta_{(2,1),(1,1),...,(1,1)}=\frac{1}{2}-\frac{2n-1}{2}\epsilon
\eeq Therefore the disorder induced interaction  ($\sim
\varepsilon\varepsilon$) is  slightly relevant: we can study
the $WD_{n}^{(p)}$ model with disorder coupled to the energy operator
and the related problem of $N$-coupled systems using the same technique as the one we
have exploited in the case of a $WD_{3}^{(p)}$ model. The analogy with
this case goes further: we show in appendix 1 that the
operator $\phi_{(1,1)(2,1)(1,1)..(1,1)}\equiv \phi$ with dimension \beq
\Delta_{\phi}\equiv\Delta_{(1,1),(2,1),(1,1),..,(1,1)}=1-2(n-1)\epsilon \eeq is, apart
from the identity, the
 only  relevant operator in the O.P.E. of
$\varepsilon(x')\varepsilon(x)$. The enlarged algebra of the fields
$\phi$ and $\varepsilon$ is given by eq.(3.8), where
$D^{\phi}_{\varepsilon\varepsilon}$ and $D^{\phi}_{\phi\phi}$ are now the
related structure constant of the $WD_{n}^{(p)}$ model. We consider
thus the action (3.3), where in this case $A_{0}^{(a)}$ describes a single
$WD_{n}^{(p)}$ model: the correspondent $\beta$-functions take the
form (3.9)
with the values of the structure constants (up to the first order in $\epsilon$)
calculated in appendix 1: \beq
D^{\phi}_{\varepsilon\varepsilon}=\sqrt{\frac{2n-1}{n}}+O(\epsilon),\quad
D^{\phi}_{\phi\phi}=\frac{4(n-1)}{\sqrt{n(2n-1)}}+O(\epsilon)
\label{structuren} \eeq In the case $N=0$, in addition to the
unstable trivial fixed point $(0,0)$, there is  one stable fixed point
\beq
g_{*}=0,\quad \lambda_{*}=2\sqrt{n(2n-1)}\epsilon \label{ap}
\eeq
For $N\geq 2$, there are two other supplementary fixed points, both
node points of the RG flow:
\begin{eqnarray}\label{Appendix-point-fix-3}
\lambda^{(1)}_{*} & = & \sqrt{n(2n-1)}\epsilon\left(1-(N-2)\sqrt{\frac{n(n-1)}{N-1+n(n-1)N^{2}}}\right)\nonumber \\
g^{(1)}_{*} & = & +2(2n-1)\epsilon\sqrt{\frac{n(n-1)}{N-1+n(n-1)N^{2}}}
\end{eqnarray}
\begin{eqnarray}\label{Appendix-point-fix-4}
\lambda^{(2)}_{*}&=&\sqrt{n(2n-1)}\epsilon\left(1+(N-2)\sqrt{\frac{n(n-1)}{N-1+n(n-1)N^{2}}}\right)\nonumber \\
g^{(2)}_{*}&=&-2(2n-1)\epsilon\sqrt{\frac{n(n-1)}{N-1+n(n-1)N^{2}}}
\end{eqnarray}
In order to compute the central charges at the fixed points for the
disorder problem and for the coupled models, we can directly apply the
result (4.9), provided the new values of the structure
constants (5.6) and of the dimension of operators (5.4) and (5.5) are
taken into account.
For the quenched problem, the central charge
$c_{dis.}^{(n)}$ at the
stable fixed point (5.7) is 
\beq
c_{dis.}^{(n)}=\lim_{N\to 0}\frac{c(0,2\sqrt{n(2n-1)}\epsilon)}{N}=c(WD_{n}^{(p)})-4n(n-1)(2n-1)\epsilon^{3}
\approx c(WD_{n}^{(p-1)})\eeq 
The correspondent field
theory is described by the $WD_{n}^{(p-1)}$model, as found already
 in [1]. 
In the coupled model, the central charge $c_{coupl.}^{(n)}$ at the same
fixed point is $c_{coupl.}^{(n)}=Nc(WD_{n}^{(p-1)})$ and the associated
critical theory is described by $N$ decoupled $WD_{n}^{(p-1)}$
models.

At the new fixed points (\ref{Appendix-point-fix-3}) and (5.9) the central 
charges $c_{1,2}^{(n)}$ turn out to be:
\beq
c_{1,2}^{(n)}=Nc(WD_{n}^{(p)})-2Nn(n-1)(2n-1)\left[1\mp(N-2)\sqrt{\frac{n(n-1)}{N-1+n(n-1)N^2}}\right]\epsilon^{3}
\eeq 
As the RG-flow behavior is the same for all $n\geq3$, the results we have shown in section 4.1 for the
disordered $WD_{3}^{(p)}$ model and for the system of $N$ coupled
$WD_{3}^{(p)}$ models are valid for the whole series of $WD_{n}^{(p)}$
conformal theories. 

Finally we give the direct  generalization of eq.(4.6):
\begin{eqnarray}
\Delta_{\varepsilon_{s}}^{(1),(2)}&=&\Delta_{\varepsilon}+(2n-1)\left(1\pm
N\sqrt{\frac{n(n-1)}{N-1+n(n-1)N^{2}}}\right)\epsilon \nonumber \\
\Delta_{\varepsilon_{a}}^{(1),(2)}&=&\Delta_{\varepsilon}+(2n-1)\left(1\mp
N\sqrt{\frac{n(n-1)}{N-1+n(n-1)N^{2}}}\right)\epsilon
\end{eqnarray}
\section{Conclusions}
As has been stressed throughout the paper, for every small value of $\epsilon$:
\beq
\epsilon=\frac{1}{p+1},\quad p\gg 1
\eeq
we have a unitary model, $WD^{(p)}_{n}$,
associated to it. When a set of $N\geq 2$
of such models are coupled and brought, by fine tuning of the couplings $g$ and
$\lambda$, to one of the two newly found fixed points, (5.8) or (5.9), the
corresponding critical theory should also be described by a unitary conformal
field theory. As we have a fixed point
for every $p$, we should have a unitary series of new conformal
theories, with $p$ being a parameter, accumulating towards $p=\infty $.
Moreover, we should have a unitary series for every $n$, of $WD^{(p)}_{n},
n=3,4,5,... $.

These new theories, presently unknown, have to incorporate into them the
permutational symmetry $S_{N}, N=2,3,...$.
This symmetry has to be incorporated
into the chiral algebra of these theories.
As the symmetry is discrete, the natural
suggestion would be that it should be represented
by parafermionic currents.

For $N\geq 3$, the group $S_{N}$ is non-Abelian. Looking at the expression of
the central charge, eq.(5.11), we observe that the correction term which
we have calculated is non-rational for $N\geq 3$.

If we accept the idea of non-Abelian parafermionic conformal theories,
we have to accept also that these theories possess an infinite series of
unitary models, labeled by $p$, according to the arguments
given above. And then the formula (5.11) indicates that the central charge
for these unitary models takes non-rational values.
This feature is unusual. 

Being more precise, the argument for rational or non-rational
values of the central charge $c$, and the dimensions of the
operators, could be given as follows.
 
If one assumes that $c$, which is a function of $p$, takes
rational values for all integer values of $p$, this
 then requires that the function $c(p)$ should have
a simple rational form, like $c(p)=Q(p)/P(p)$, where
$Q(p),P(p)$ are two polynomials of $p$ with rational
valued coefficients. If the function $c(p)$ of such a
form is developped in a series of $1/p$, the coefficients
will all be  rational.

Still, it should be admitted that we don't know in
fact if the perturbative expansion in $\epsilon\sim
1/p$, which we define in this paper, represents a
convergent series. The series might also be asymptotic,
as it is often the case in perturbative expansions in
field theory. This then allows for two possibilities.
 
The first possibility will be that the series is
convergent, representing an analytic function $c(p)$,
like this is two case for the perturbed minimal
models in [5].
 
In this case, according to the argument given above,
one could actually judge on rationality-
non rationality with the coefficients of the expansion.

Saying it again, for having rational values of $c(p)$,
for all integer $p$, a complicated function
$c(p)$ will not be allowed.
 
Then the coefficients of the expansion will also have to be rational.
 
The second possibility will be that the series in
$\epsilon$ is not convergent, that it is only
asymptotic. In this case our arguments do not
apply and the exact values of the central charge
might well be rational, in spite of the irrational
coefficients of the expansion.
 
In view of this second possibility, our conclusions
on non-rationality should not be viewed as definite.
 
Construction of the corresponding exact conformal
field theories represents a theoretical challenge.

\vspace{3cm}

{\bf Acknowledgements}

Discussions with D. Bernard, J. Jacobsen, A. LeClair, M. Picco,  have
been stimulating.
\newpage
\appendix
\section{Appendix 1}
\subsection{Relevant Operators in the O.P.E. of $\varepsilon(x')\varepsilon(x)$}

In section 3 we have explained how to deduce the O.P.E. of
$\varepsilon(x')\varepsilon(x)$ from the integral representation of
the energy four-point correlation function: within the set of vectors
\{$2\vec{\beta}_{\varepsilon}+\alpha_{-}\sum_{a=1}^{3}q_{a}\vec{e}_{a}$\}
 where $q_{1}=0,1,2$, $q_{2}=0,1$, and $q_{3}=0,1$, we search for the
ones which decompose into positive integer numbers of
\{$\alpha_{-}\frac{\vec{\omega}_{a}}{2}$\}, basic vectors of the
representation lattice, i.e.: 
\beq
2\vec{\beta}_{\varepsilon}+\alpha_{-}\sum_{a=1}^{3}q_{a}\vec{e}_{a}=\alpha_{-}\sum_{a=1}^{3}\frac{1-n_{a}'}{2}\vec{\omega}_{a}
\eeq
with $n_{a}'\geq 1$. In this case the vertex operator
$V_{2\vec{\beta}_{\varepsilon}+\alpha_{-}\sum_{a=1}^{3}q_{a}\vec{e}_{a}}$
 belongs to the physical sector of the Kac table and the correspondent
primary operator $\phi_{(n_{1}',1)(n_{2}',1)(n_{3}',1)}$ appears in
the energy algebra.
Using eq.(2.19) and eq.(2.20), the condition (A.1) is equivalent to:
\begin{eqnarray}
n_{1}'&=& 3-2q_{1}+q_{3}+q_{2} \nonumber \\
n_{2}'&=& 1 -2q_{2}+q_{1} \nonumber \\
n_{3}'&=& 1-2q_{3}+q_{1}
\end{eqnarray}
Testing all the possible set of values ($q_{1},q_{2},q_{3}$), it's
easy to see that the system (A.2) admits only three solutions; they
correspond to the identity operator (for $q_{1}=2, q_{3}=1, q_{2}=1$),
to the operator $\phi$ (for $q_{1}=1, q_{3}=0, q_{2}=0$) and to the
operator 
$\phi_{(3,1)(1,1)(1,1)}$ (for $q_{1}=q_{2}=q_{3}=0$). According to
 eq.(2.31)  $\Delta_{(3,1)(1,1)(1,1)}=2$: the only relevant operators in
the energy algebra are the identity and $\phi$.

Now, in order to verify that the algebra formed by these two fields
doesn't contain any other relevant operator, we have to check the
O.P.E. of $\phi(x')\phi(x)$. Following the same procedure used before,
it's easy to verify that, apart from identity, the $\phi$ field is the only
relevant operator in the $\phi$-algebra, as already discussed in [1].
We consider in this case the integral representation of the four-point correlation function
$G'(x)\equiv\langle\phi(0)\phi(x)\phi(1)\phi(\infty)\rangle$:
\beq G'(x)\propto \int..\int\langle
V_{\vec{\beta_{\phi}}}(0)V_{\vec{\beta_{\phi}}}(x)V_{\vec{\beta_{\phi}}}(1)
V_{\vec{\beta_{\bar{\phi}}}}(\infty)V_{1}^{-}(\mu_{1})V_{1}^{-}(\mu_{2})V_{2}^{-}(\xi_{1})V_{2}^{-}(\xi_{2})
V_{3}^{-}(\nu_{1})V_{3}^{-}(\nu_{2})
\rangle
\eeq
where the variables $\mu_{1},\mu_{2},\xi_{1},\xi_{2},\nu_{1},\nu_{2}$
are integrated over the 2D plane. 
The charge neutrality is satisfied according to
$-2\vec{\beta}_{\phi}+2\alpha_{-}(\vec{e}_{1}+\vec{e}_{2}+\vec{e}_{3})=0$.
 As discussed above all the operators
$\phi_{(n_{1}',1)(n_{2}',1)(n_{3}',1)}$
 present in the
 O.P.E. of $\phi(x')\phi(x)$ are given by the condition
\beq
2\vec{\beta}_{\phi}+\alpha_{-}\sum_{a=1}^{3}q_{a}\vec{e}_{a}=\alpha_{-}\sum_{a=1}^{3}\frac{1-n_{a}'}{2}\vec{\omega}_{a}
\eeq
where  $q_{1},q_{2},q_{3}$ can assume the values $0,1,2$; eq. (A.4) is
equivalent to:
\begin{eqnarray}
n_{1}'&=& 1-2q_{1}+q_{3}+q_{2} \nonumber \\
n_{2}'&=& 3 -2q_{2}+q_{1} \nonumber \\
n_{3}'&=& 3+q_{1}-2q_{3}
\end{eqnarray} 
By eq.(A.5) we have determined, apart from the identity, four primary
operators: $\phi_{(1,1)(3,1)(3,1)}$, $\phi_{(2,1)(3,1)(1,1)}$,
$\phi_{(2,1)(1,1)(3,1)}$ and $\phi$;
the only field whose dimension is smaller than the unity is $\phi$.
\subsection{Relevant Operators in the O.P.E. of
$\varepsilon(x')\varepsilon(x)$ for the $WD_{n}^{(p)}$ model with $n\geq4$}
As already discussed  in section 5, the energy operator $\varepsilon$
of the $WD_{n}^{(p)}$ model is associated to the primary operator
$\phi_{(2,1)(1,1)..(1,1)}$. We show below that in the correspondent algebra there is,
apart from the identity, only
one relevant primary operator, namely
$\phi_{(1,1)(2,1)(1,1)\cdots(1,1)}\equiv \phi$. In the following we define for simplicity
 $\vec{\beta}_{(2,1)(1,1)\cdots(1,1)} \equiv \vec{\beta}_{\varepsilon}$,
$2\vec \alpha_{0}-\vec \beta_{(2,1)(1,1)\cdots(1,1)}\equiv
\vec{\beta}_{\bar{\varepsilon}}$,
$\vec{\beta}_{(1,1)(2,1)(1,1)\cdots(1,1)}\equiv \vec{\beta}_{\phi}$, and
$2\vec \alpha_{0}-\vec \beta_{(1,1)(2,1)(1,1)\cdots(1,1)}\equiv
\vec{\beta}_{\bar{\phi}}$, The integral representation of the
 correspondent four-point correlation function
$G(x)\equiv\langle\varepsilon(0)\varepsilon(x)\varepsilon(1)\varepsilon(\infty)\rangle$
is:
\beq
G(x)\propto \int \cdots \int\langle
V_{\vec{\beta_{\varepsilon}}}(0)V_{\vec{\beta_{\varepsilon}}}(x)V_{\vec{\beta_{\varepsilon}}}(1)
V_{\vec{\beta_{\bar{\varepsilon}}}}(\infty)\underbrace{V_{1}^{-}\cdots V_{1}^{-}}_{m_{1}
times}\underbrace{V_{2}^{-}\cdots V_{2}^{-}}_{m_{2}
times}\cdots\underbrace{V_{n}^{-}\cdots V_{n}^{-}}_{m_{n}
times}\rangle
\eeq
 where the screenings $V_{a}^{-}$ with $a=1,\cdots,n$ are integrated
over the 2D plane. Using the quadratic form (5.2) it's easy to see
that  the charge neutrality condition 
\beq
-2\vec{\beta}_{\varepsilon}+\alpha_{-}\sum_{a}^{n}m_{a}\vec{e}_{a}=0
\eeq
imposes  $m_{a}=2$ for $a=1,\cdots,n-2$ and
$m_{n-1}=m_{n}=1$.
For the same arguments discussed in the  previous section, we search for the
vectors $\vec{\beta}_{(n_{1}',1)(n_{2}',1)\cdots(n_{n}',1))}$ with
$n_{a}'\geq 1$ (see 5.1) such that:
\beq
2\vec{\beta}_{\varepsilon}+\alpha_{-}\sum_{a=1}^{n}q_{a}\vec{e}_{a}=\vec{\beta}_{(n_{1}',1)(n_{2}',1)\cdots(n_{n}',1)}
\eeq 
where $q_{a}=0,1,2$ for $a=1,\cdots, n-2$, $q_{n-1}=0,1$ and $q_{n}=0,1$.
Using eq.(5.2), the condition (A.8) is equivalent to the following
system of equations:
\begin{eqnarray}
&&n_{1}'=3-2q_{1}+q_{2} \nonumber \\
&&n_{a}'=1-2q_{a}+q_{a-1}+q_{a+1} \quad 2\leq a<n-2 \nonumber \\
&&n_{n-2}'=1-2q_{n-2}+q_{n-3}+q_{n-1}+q_{n} \nonumber \\
&&n_{n-1}'=1-2q_{n-1}+q_{n-2}\nonumber \\
&&n_{n}'=1-2q_{n}+q_{n-2}
\end{eqnarray}
Eq.(A.9) gives us all the possible primary operators
$\phi_{(n_{1}',1)(n_{2}',1)\cdots(n_{n}',1)}$ present in the energy algebra.
Once again, testing all the possible set of values
($q_{1},\cdots,q_{n}$) it's easy to verify that, apart from the
identity, there are only two  solutions: they 
 correspond to the operator $\phi_{(3,1)(1,1)\cdots(1,1)}$ (for $q_{a}=0$), with
dimension greater than unity, and to the operator
$\phi_{(1,1)(2,1)(1,1)\cdots(1,1)}\equiv \phi$ (for $q_{1}=1$ and $q_{a}=0,\;a\geq
2$), whose  dimension is given by eq(5.5). In the operator product
decomposition 
$\phi(x')\phi(x)$,  we have checked that the only relevant field is
$\phi$ itself: by studying the correspondent
four-point correlation function we obtain the following system:
\begin{eqnarray}
&&n_{1}'= 1-2q_{1}+q_{2} \nonumber \\  
&&n_{2}'=3-2q_{2}+q_{1}+q_{3}\nonumber \\
&&n_{i}'=1-2q_{i}+q_{i-1}+q_{i+1}\quad  3\leq i\leq n-3 \nonumber \\
&&n_{n-2}'=1-2q_{n-2}+q_{n-3}+q_{n-1}+q_{n} \nonumber \\
&&n_{n-1}'=1-2q_{n-1}+q_{n-2}\nonumber \\
&&n_{n}'=1-2q_{n}+q_{n-2}
\end{eqnarray}
with the integers $q_{1},q_{n-1},q_{n}=0,1,2$ and $q_{a}=0,1,2,3,4$ for
$a=2,\cdots,n-2$. According to eq.(A.10), the $\phi$ field is the
only operator with dimension smaller than unity which appears in the $\phi$-algebra. 
\section{Computation of the structure constants
$D_{\varepsilon\varepsilon}^{\phi}$ and $D_{\phi\phi}^{\phi}$ for the
$WD_{n}^{(p)}$ models}
The structure constants  $D_{\varepsilon\varepsilon}^{\phi}$ and
$D_{\phi\phi}^{\phi}$ are determined by two three-point correlation
functions of the unperturbed $WD_{n}^{(p)}$ model:
\begin{eqnarray}
&&D_{\varepsilon\varepsilon}^{\phi}=\langle\varepsilon(0)\varepsilon(1)\phi(\infty)\rangle
\nonumber \\
&&D_{\phi\phi}^{\phi}=\langle\phi(0)\phi(1)\phi(\infty)\rangle 
\end{eqnarray}
where $\phi$ corresponds to the primary operator
$\phi_{(1,1)(2,1)(2,1)}$ for the $WD_{3}^{(p)}$ model and to the
primary operator $\phi_{(1,1)(2,1)(1,1)\cdots(1,1)}$ for the
$WD_{n}^{(p)}$ model with $n\geq4$. In order to compute their values using the Coulomb-Gas representation,
we have to take into account that the vertex operators
$V_{\vec{\beta}_{\varepsilon}}$,
$V_{\vec{\beta}_{\bar{\varepsilon}}}$, $V_{\vec{\beta}_{\phi}}$ and
$V_{\vec{\beta}_{\bar{\phi}}}$ can acquire non-trivial normalization
factors  $N_{\varepsilon}$,
$N_{\bar{\varepsilon}}$, $N_{\phi}$ and $N_{\bar{\phi}}$ \cite{DotsenkoFateev} such that:
\begin{eqnarray}
&&\varepsilon(x)=N_{\varepsilon}^{-1}V_{\vec{\beta}_{\varepsilon}}(x)=N_{\bar{\varepsilon}}^{-1}
V_{\vec{\beta}_{\bar{\varepsilon}}}(x) \nonumber \\
&&\phi(x)=N_{\phi}^{-1}V_{\vec{\beta}_{\phi}}(x)=N_{\bar{\phi}}^{-1}V_{\vec{\beta}_{\bar{\phi}}}(x)
\end{eqnarray}
After imposing the  fields normalizations
$\langle\varepsilon(1)\varepsilon(0)\rangle=N_{\varepsilon}N_{\bar{\varepsilon}}=1$
and
$\langle\phi(1)\phi(0)\rangle=N_{\phi}N_{\bar{\phi}}=1$, 
 we need to compute four integrals:
\begin{eqnarray}
&&I_{1}\equiv \prod_{i}^{n}\frac{(-)^{m_{i}^{(1)}}}{m_{i}^{(1)}!}\int..\int\langle
V_{\vec{\beta}_{\varepsilon}}(0)
V_{\vec{\beta}_{\phi}}(1)V_{\vec{\beta}_{\bar{\varepsilon}}}(\infty)\underbrace{V_{1}\cdots
V_{1}}_{m_{1}^{(1)}\;times}\cdots
\underbrace{V_{n}\cdots
V_{n}}_{m_{n}^{(1)}\;times}\rangle=N_{\phi}D_{\varepsilon\varepsilon}^{\phi} \\
&&I_{2}\!\equiv\! \prod_{i}^{n}\frac{(-)^{m_{i}^{(2)}}}{m_{i}^{(2)}!}\int..\int\langle
V_{\vec{\beta}_{\varepsilon}}(0)
V_{\vec{\beta}_{\phi}}(1)V_{\vec{\beta}_{\bar{\phi}}}(\infty)\underbrace{V_{1}\cdots
V_{1}}_{m_{1}^{(2)}\;times}\cdots
\underbrace{V_{n}\cdots
V_{n}}_{m_{n}^{(2)}\;times}\rangle\!=\!N_{\varepsilon}^{2}N_{\phi}^{-1}D_{\varepsilon\varepsilon}^{\phi} \\
&&I_{3}\equiv \prod_{i}^{n}\frac{(-)^{m_{i}^{(3)}}}{m_{i}^{(3)}!}\int..\int\langle
V_{\vec{\beta}_{\phi}}(0)
V_{\vec{\beta}_{\phi}}(1)V_{\vec{\beta}_{\bar{\phi}}}(\infty)\underbrace{V_{1}\cdots
V_{1}}_{m_{1}^{(3)}\;times}\cdots
\underbrace{V_{n}\cdots
V_{n}}_{m_{n}^{(3)}\;times}\rangle=N_{\phi}D_{\phi
\phi}^{\phi} \\
&&I_{4}\equiv \prod_{i}^{n}\frac{(-)^{m_{i}^{(4)}}}{m_{i}^{(4)}!}\int..\int\langle
V_{\vec{\beta}_{\varepsilon}}(0)
V_{\vec{\beta}_{\varepsilon}}(1)V_{\vec{\beta}_{2\vec{\alpha}_{0}}}(\infty)\underbrace{V_{1}\cdots
V_{1}}_{m_{1}^{(4)}\;times}\cdots
\underbrace{V_{n}\cdots
V_{n}}_{m_{n}^{(4)}\;times}\rangle=N_{\varepsilon}^{2}
\end{eqnarray}
where the integration over the 2D plane of the screenings, whose number is fixed by the charge neutrality condition, is intended.
In eq. (B.6) the vertex $V_{\vec{\beta}_{2\vec{\alpha}_{0}}}$
corresponds to the identity operator $I$ and the relative normalization
constant $N_{\vec{\beta}_{2\vec{\alpha}_{0}}}=N_{I}=1$. According to the general form of the correlation function of $N$
vertex operator $V_{\vec{\beta}_{i}}$:
\beq
\langle V_{\vec{\beta}_{1}}(\xi_{1})\cdots
V_{\vec{\beta}_{N}}(\xi_{N})\rangle=\prod_{i<j}|\xi_{i}-\xi_{j}|^{4\vec{\beta}_{i}\vec{\beta}_{j}},
\eeq
and using the formula  
\begin{eqnarray}\label{Appendix-C-integral}
C(\alpha,\beta)&=&\int d^{2}\xi|\xi|^{2\alpha}|\xi-1|^{2\beta}\cr
&=&\pi\frac{\Gamma(1+\alpha)\Gamma(1+\beta)\Gamma(-1-\alpha-\beta)}{\Gamma(-\alpha)\Gamma(-\beta)\Gamma(2+\alpha+\beta)}
\end{eqnarray}
and
\begin{eqnarray}\label{Appendix-K-integral}
K(\alpha,\rho)&=&\int d^{2}\xi
d^{2}\zeta|\xi|^{2\alpha}|\xi-1|^{2\alpha}|\zeta|^{2\alpha}|\zeta-1|^{2\alpha}|\zeta-\xi|^{4\rho}\cr
&=&2\pi^{2}\frac{\Gamma(2\rho)\Gamma(1-\rho)}{\Gamma(\rho)\Gamma(1-2\rho)}\prod_{i=0}^{1}\frac{\Gamma^{2}(1+\alpha+i\rho)\Gamma(-1-2\alpha-(1+i)\rho)}{\Gamma^{2}(-\alpha-i\rho)\Gamma(2+2\alpha+(1+i)\rho)}
\end{eqnarray}
where $\Gamma(x)$ is the Gamma-function, the integrals (B.3)-(B.6) have been
determined. Eq.(B.8) and eq.(B.9) represent special cases of a family of integrals
calculated in \cite{DotsenkoFateev}.  

In terms of these integrals, the structure constants are :
\begin{eqnarray}
&&D_{\phi\phi}^{\phi}=I_{3}\sqrt{\frac{I_{2}}{I_{1}I_{4}}}\nonumber \\
&&D_{\varepsilon\varepsilon}^{\phi}=\sqrt{\frac{I_{1}I_{2}}{I_{4}}}
\end{eqnarray}
\subsection{Computation of $I_{1}$}
The charge neutrality condition for $I_{1}$ is:
\beq
-\vec{\beta}_{\phi}+\alpha_{-}\sum_{a=1}^{n}m_{a}^{(1)}\vec{e}_{a}=0
\eeq
It is satisfied for $m_{1}^{(1)}=m_{n-1}^{(1)}=m_{n}^{(1)}=1$,
$m_{a}^{(1)}=2,\: a=2,\cdots,n-2$. Using eq.(B.7) and taking into
account all the scalar products, easily computed from the quadratic
form (5.2), the
integral $I_{1}$ takes the form:
\begin{eqnarray}
I_{1}&=&-2^{3-n}\int \cdots \int d^{2}\xi_{1}\prod_{a=2}^{n-2} \prod_{k=1}^{2}
d^{2}\xi_{a}^{(k)}
d^{2}\xi_{n-1}d^{2}\xi_{n}\times \nonumber \\
 &&
\times |\xi_{1}|^{-2\alpha_{-}^{2}}\prod_{k=1}^{2}|\xi_{2}^{(k)}-\xi_{1}|^{-2\alpha_{-}^{2}}|\xi_{2}^{(k)}-1|^{-2\alpha_{-}^{2}}
\prod_{a=2}^{n-3}\prod_{k,l=1}^{2}|\xi_{a}^{(k)}-\xi_{a+1}^{(l)}|^{-2\alpha_{-}^{2}}\times
\nonumber \\
&&\times
\prod_{a=2}^{n-2}|\xi_{a}^{(1)}-\xi_{a}^{(2)}|^{4\alpha^{2}_{-}}\prod_{k=1}^{2}|\xi_{n-2}^{(k)}-\xi_{n-1}|^{-2\alpha_{-}^{2}}
|\xi_{n-2}^{(k)}-\xi_{n}|^{-2\alpha_{-}^{2}}
\end{eqnarray}
where $\alpha_{-}^{2}=p/(p+1)\approx 1-\epsilon$.  Integrating over
the variables  $\xi_{n}$, and $\xi_{n-1}$ we obtain by eq. (B.8):
\begin{eqnarray}
I_{1}&=&-2^{3-n}C^{2}(-\alpha_{-}^{2},-\alpha^{2}_{-})\int \cdots \int d^{2}\xi_{1}\prod_{a=1}^{n-2} \prod_{k=1}^{2}
d^{2}\xi_{a}^{(k)}\times \nonumber \\
 &&
\times |\xi_{1}|^{-2\alpha_{-}^{2}}\prod_{k=1}^{2}|\xi_{2}^{(k)}-\xi_{1}|^{-2\alpha_{-}^{2}}|\xi_{2}^{(k)}-1|^{-2\alpha_{-}^{2}}
\prod_{a=2}^{n-3}\prod_{k,l=1}^{2}|\xi_{a}^{(k)}-\xi_{a+1}^{(l)}|^{-2\alpha_{-}^{2}}\times
\nonumber \\
&&\times
\prod_{a=2}^{n-3}|\xi_{a}^{(1)}-\xi_{a}^{(2)}|^{4\alpha^{2}_{-}}|\xi_{n-2}^{(1)}-\xi_{n-2}^{(2)}|^{4-4\alpha^{2}_{-}}
\end{eqnarray}
Then using eq. (B.9) we integrate in order  over the couple of variables 
$\xi_{n-2}^{(k)},\cdots,\xi_{2}^{(k)}$ and over the
variable $\xi_{1}$; the result is: 
\begin{eqnarray}
I_{1}&=&-2^{3-n}C^{2}(-\alpha^{2}_{-},-\alpha^{2}_{-})C(-\alpha^{2}_{-},2\alpha^{2}_{-}+2n)\prod_{i=1}^{n-3}
K(-\alpha^{2}_{-},i(1-\alpha^{2}_{-}))
\end{eqnarray}
\subsection{Computation of $I_{2}$}
According to :
\beq
-2\vec{\beta}_{\varepsilon}+\vec{\beta}_{\phi}+\alpha_{-}\vec{e_{1}}=0
\eeq
we have $m_{1}^{(2)}=1$ and $m_{a}^{(2)}=0$ for $a=2,\cdots,n$; the integral (B.4) reads:
\begin{eqnarray}
I_{2}&=&-\int d^{2}\xi_{1}
\langle V_{\vec{\beta_{\varepsilon}}}(0)V_{\vec{\beta_{\varepsilon}}}(1)V_{\vec{\beta_{\bar{\phi}}}}(\infty)V^{-}_{1}(\xi_{1})
\rangle \nonumber \\ 
&=&-\int d^{2}\xi_{1}|\xi|^{-2\alpha^{2}_{-}}|\xi_{1}-1|^{-2\alpha_{-}^{2}}\nonumber
\\
&=& -C(-\alpha^{2}_{-}, -\alpha^{2}_{-})
\end{eqnarray}
\subsection{Computation of $I_{3}$}
The integral (B.5) must satisfy the charge neutrality condition
(B.11); it takes the form:
\begin{eqnarray}
I_{3}&=&-2^{3-n}\int \cdots \int d^{2}\xi_{1}\prod_{a=2}^{n-2} \prod_{k=1}^{2}
d^{2}\xi_{a}^{(k)}
d^{2}\xi_{n-1}d^{2}\xi_{n}\times \nonumber \\
 &&
\times \prod_{k=1}^{2}|\xi_{2}^{(k)}-\xi_{1}|^{-2\alpha_{-}^{2}}|\xi_{2}^{(k)}-1|^{-2\alpha_{-}^{2}}|\xi_{2}^{(k)}|^{-2\alpha_{-}^{2}}
\prod_{a=2}^{n-3}\prod_{k,l=1}^{2}|\xi_{a}^{(k)}-\xi_{a+1}^{(l)}|^{-2\alpha_{-}^{2}}
\times
\nonumber \\
&&\times
\prod_{a=2}^{n-2}|\xi_{a}^{(1)}-\xi_{a}^{(2)}|^{4\alpha^{2}_{-}}\prod_{k=1}^{2}|\xi_{n-2}^{(k)}-\xi_{n-1}|^{-2\alpha_{-}^{2}}
|\xi_{n-2}^{(k)}-\xi_{n}|^{-2\alpha_{-}^{2}}
\end{eqnarray}
Following the same order of integrations used in the computation of
 the integral (B.3), with the difference that we integrate first over the variable
 $\xi_{1}$ and then over the couple of variables
 $\xi_{2}^{(1)},\xi_{2}^{(2)}$, we obtain:
 \beq
I_{3}=-2^{3-n}C^{3}(-\alpha^{2}_{-},-\alpha^{2}_{-})K(-\alpha^{2}_{-},\frac{2n-5}{2}-(n-2)\alpha^{2}_{-})\prod_{i=1}^{n-4}
K(-\alpha^{2}_{-},i(1-\alpha^{2}_{-}))
\eeq
\subsection{Computation of $I_{4}$}
According to the charge neutrality condition:
\beq
-2\vec{\beta}_{\varepsilon}+\sum_{a=1}^{n}m_{a}^{(4)}\vec{e}_{a}=0
\eeq
we have $m_{a}^{(4)}=2$ for $a=1,\cdots,n-2$,
$m_{n-1}^{(4)}=m_{n}^{(4)}=1$. The integral (B.6) has the form:
\begin{eqnarray}
I_{4}&=&2^{2-n}\int \cdots \int \prod_{a=1}^{n-2} \prod_{k=1}^{2}
d^{2}\xi_{a}^{(k)}
d^{2}\xi_{n-1}d^{2}\xi_{n}\times \nonumber \\
 &&
\times \prod_{k=1}^{2}|\xi_{1}^{(k)}-1|^{-2\alpha_{-}^{2}}|\xi_{1}^{(k)}|^{-2\alpha_{-}^{2}}
\prod_{a=1}^{n-3}\prod_{k,l=1}^{2}|\xi_{a}^{(k)}-\xi_{a+1}^{(l)}|^{-2\alpha_{-}^{2}}
\times
\nonumber \\
&&\times
\prod_{a=1}^{n-2}|\xi_{a}^{(1)}-\xi_{a}^{(2)}|^{4\alpha^{2}_{-}}\prod_{k=1}^{2}|\xi_{n-2}^{(k)}-\xi_{n-1}|^{-2\alpha_{-}^{2}}
|\xi_{n-2}^{(k)}-\xi_{n}|^{-2\alpha_{-}^{2}}
\end{eqnarray}
Integrating over the variables
$\xi_{n},\xi_{n-1},\xi_{n-2}^{(k)},\cdots,\xi_{1}^{(k)}$, the integral
(B.6) assumes the value:
\beq
I_{4}=C^{2}(-\alpha^{2}_{-},-\alpha^{2}_{-})\prod_{i=1}^{n-2}K(-\alpha^{2}_{-},i(1-\alpha^{2}_{-}))
\eeq
Developing in $\epsilon$ the integrals we have computed and using eq.
(B.10), the eq.(5.6) are obtained. 
\section{RG Equations and Renormalization of energy operators}
Using the O.P.E.(3.8) and the dimensions of the energy and $\phi$ fields (eq.(5.4) and (5.5)), the operator algebra of the
perturbing terms is:
\begin{eqnarray}
\sum_{a\neq b}^{N}\:\left(\varepsilon_{a}\varepsilon_{b}\right)(x)
\sum_{c\neq d}^{N}\:\left(\varepsilon_{c}\varepsilon_{d}\right)(y)&\to&
4(N-2)|x-y|^{-2+2(2n-1)\epsilon}\sum_{a\neq b}^{N}\:\left(\varepsilon_{a}\varepsilon_{b}\right)(y)+ \nonumber \\
&&+4(N-1)D_{\varepsilon\varepsilon}^{\phi}|x-y|^{-2+4n}\sum_{a=1}^{N}\phi_{a}(y)+\cdots
\nonumber \\
\!\!\sum_{a\neq
b}^{N}\:\left(\varepsilon_{a}\varepsilon_{b}\right)(x)\sum_{a=1}^{N}\phi_{a}(y)\:\:\to&&\!\!\!\!\!\!\!\!\!\!
2D_{\varepsilon\varepsilon}^{\phi}|x-y|^{-2+2(n-1)}
\sum_{a\neq
b}^{N}\:\left(\varepsilon_{a}\varepsilon_{b}\right)(y)+\cdots\nonumber \\
\sum_{a=1}^{N}\phi_{a}(x)\:\sum_{b=1}^{N}\phi_{a}(y)\quad\to\quad\:&&\!\!\!\!\!\!\!\!\!\!\!\!\!\!\!\!\!D_{\phi\phi}^{\phi}|x-y|^{-2+2(n-1)}\sum_{a=1}^{N}\phi_{a}(y)+\cdots
\end{eqnarray}
where we have omitted the irrelevant terms. By eq.(C.1) the 1-loop
RG-equations can be easily obtained (see for example \cite{dpp1} or \cite{Bernard}):
\begin{eqnarray}
         \beta_{g}&=&2(2n-1)\epsilon
         g-4\pi(N-2)g^{2}-4\pi D_{\varepsilon\varepsilon}^{\phi}\lambda g \nonumber\\
         \beta_{\lambda}&=&
        2(n-1)\epsilon\lambda-\pi D_{\phi\phi}^{\phi}
       \lambda^{2}-2\pi(N-1)D_{\varepsilon\varepsilon}^{\phi}g^{2}
\end{eqnarray}
With the redefinitions $g\to g/(4\pi)$ and $\lambda\to\lambda/(2\pi)$,
we find the eq.(3.9).

Similarly the renormalized energy operators $\varepsilon_{c}'(x)$ (3.11)
 can be computed at the first
order \cite{dpp1} from the following operator product
decomposition
\begin{eqnarray}
\sum_{a\neq
b}^{N}\:\left(\varepsilon_{a}\varepsilon_{b}\right)(x)\varepsilon_{c}(y)&\to&
2|x-y|^{-2+2(2n-1)\epsilon}
\sum_{a\neq c}^{N}\varepsilon_{a}(y)+\cdots \nonumber \\
\sum_{a}^{N}\phi_{a}(x)\varepsilon_{c}(y)\:\to&&\!\!\!\!\!\!\!
D_{\varepsilon\varepsilon}^{\phi}\:|x-y|^{-2+2(n-1)\epsilon}\varepsilon_{c}(y)+\cdots
\end{eqnarray}
\newpage
\begin{figure}
\begin{center}
\epsfxsize=400pt{\epsffile{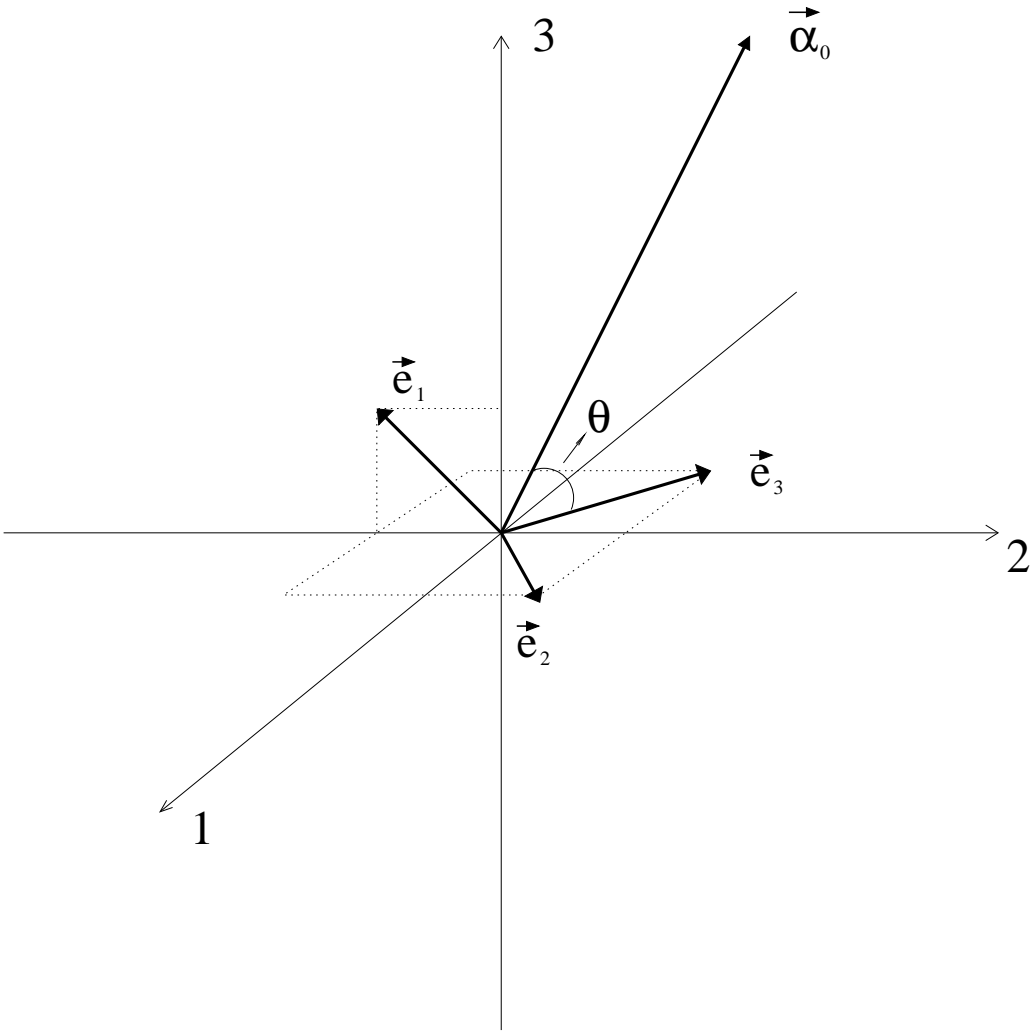}}
\end{center}
\protect\caption[2]{\label{pd} Screening geometry of the
$WD_{3}^{(p)}$ model.} 
\end{figure}    
\begin{figure}
\begin{center}
\epsfxsize=400pt{\epsffile{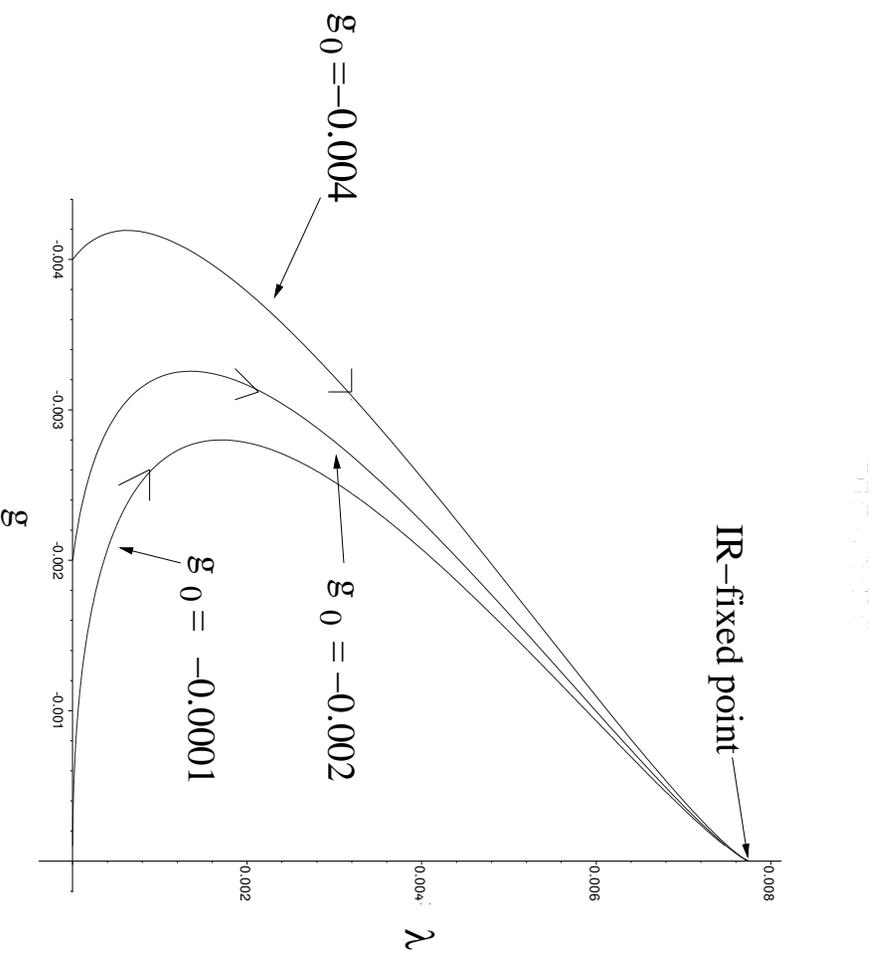}}
\end{center}
\protect\caption[2]{\label{pe} RG-flow of the $WD_{3}^{(p)}$ model
with disorder.} 
\end{figure}
\begin{figure}
\begin{center}
\epsfxsize=450pt{\epsffile{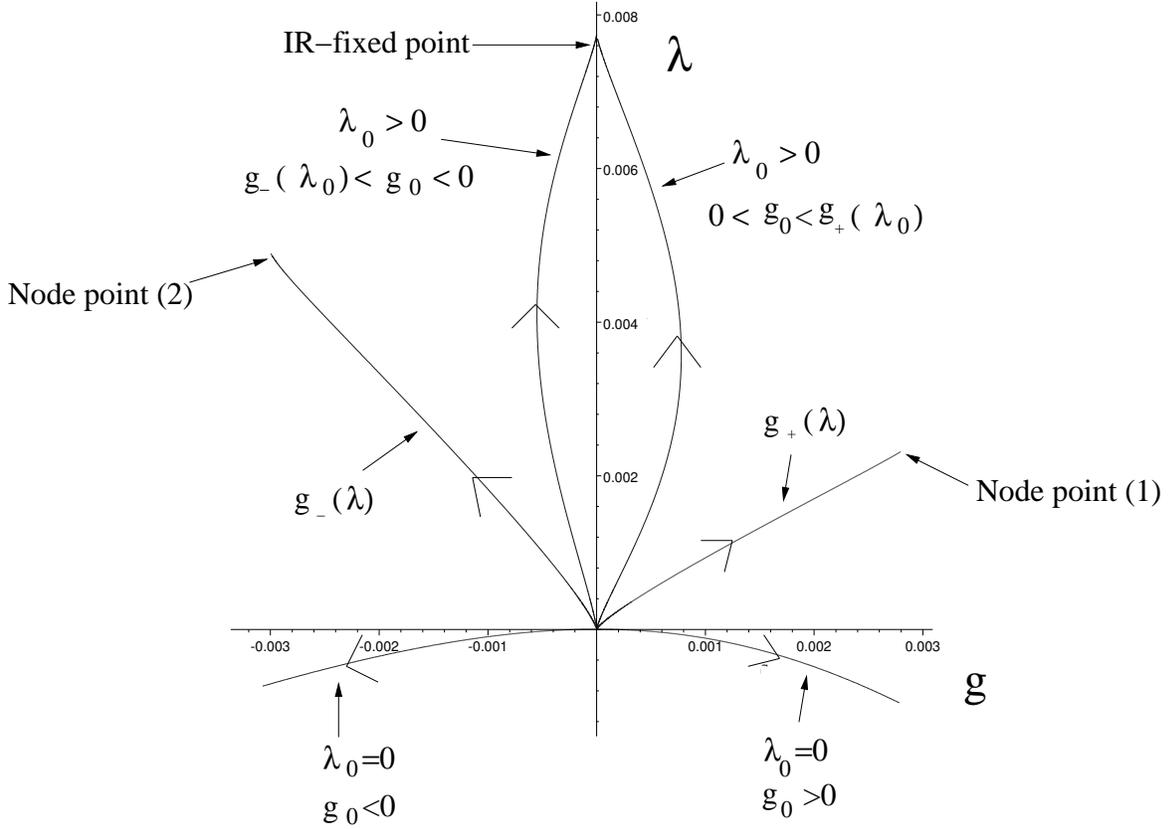}}
\end{center}
\protect\caption[2]{\label{pf} RG-trajectories of $N=3$ coupled
$WD_{3}^{(p)}$ models.} 
\end{figure}

\end{document}